\documentclass[epj]{svjour}

\usepackage{graphics}

\usepackage{amsmath}
\usepackage{amsfonts}
\usepackage{amssymb}
\usepackage{graphicx}
\usepackage{hyperref}
\usepackage{mathtools}
\usepackage{subfigure}
\usepackage{booktabs,dcolumn,caption}
\newcolumntype{d}[1]{D{.}{.}{#1}} 
\usepackage{siunitx}
\DeclareUnicodeCharacter{2212}{-}

\usepackage{lipsum}
\usepackage{mathtools}
\usepackage{cuted}

\begin{document}

\title{Red and blue shift in spherical and axisymmetric spacetimes and astrophysical constraints}
\author{Roberto Giamb\`o\inst{1,2}\thanks{roberto.giambo@unicam.it} \and Orlando Luongo\inst{1,3,4}\thanks{orlando.luongo@unicam.it} \and Lorenza Mauro\inst{5,6}\thanks{lorenza.mauro@lnf.infn.it}}

%
%
%
\institute{Scuola di Scienze e Tecnologie, Universit\`a di Camerino, Via Madonna delle Carceri 9, 62032 Camerino, Italy. \and Istituto Nazionale di Fisica Nucleare, Sezione di Perugia, Via Alessandro Pascoli 23c, 06123 Perugia, Italy. \and Dipartimento di Matematica, Universit\`a di Pisa, Largo B. Pontecorvo 5, Pisa, 56127, Italy. \and Institute of Experimental and Theoretical Physics, Al-Farabi Kazakh National University, Almaty 050040, Kazakhstan. \and Dipartimento di Matematica e Fisica, Universit\`{a} degli Studi ``Roma Tre'', Via della Vasca Navale, Roma, Italy. \and Istituto Nazionale di Fisica Nucleare (INFN), Laboratori Nazionali di Frascati, 00044 Frascati, Italy.}
\date{Received: date / Revised version: date}
%
\abstract{
We compute the red and blue shifts for astrophysical and cosmological sources. In particular, we consider low, intermediate and high gravitational energy domains. Thereby, we handle the binary system Earth - Mars as low energy landscape whereas white dwarfs and neutron stars as higher energy sources. To this end, we take into account a spherical Schwarzschild - de Sitter spacetime and an axially symmetric Zipoy - Voorhees metric  to model all the aforementioned systems. Feasible outcomes come from modelling neutron stars and white dwarfs with the Zipoy - Voorhees metric, where quadrupole effects are relevant, and framing solar system objects using a Schwarzschild - de Sitter spacetime. In the first case, large $\delta$ parameters seem to be favorite, leading to acceptable bounds mainly for neutron stars. In the second case, we demonstrate incompatible red and blue shifts with respect to lunar and satellite laser ranging expectations, once the cosmological constant is taken to  Planck satellite's best fit. To heal this issue, we suggest coarse-grained experimental setups and propose Phobos for working out satellite laser ranging in order to get more suitable red and blue shift intervals, possibly more compatible than  current experimental bounds. Implications to cosmological tensions are also debated.
\PACS{
      {04.20.−q}{Red and blue shifts. Spherical and axially - symmetric spacetimes. Compact objects. Cosmological constant.}   \and
      {04.80.Cc}{Laser Ranging.}
     } 
} 
\maketitle
\section{Introduction}

Reconciling low with high gravitational energy scales remains a subtle
issue that is so far not fully-explored by merely looking at current astrophysical data \cite{referenza1}. Thereby, probing general relativity at both short and large distances with arbitrary accuracy is seemly for guaranteeing the validity of general relativity at different energy scales \cite{ref1,ref2}.
\noindent
For instance, at primordial times the increase of gravitational energy could break down general relativity predictability,  requiring the existence of quantum gravity \cite{referenza2,referenza2a,referenza2b}. No conclusive approaches to quantum gravity exist, challenging the standard puzzle of unifying fundamental forces into a single scheme \cite{referenza2bis1,referenza2bis1a,referenza2bis2,referenza2bis3}. On the other hand, at infrared energy domains unknown ingredients, namely dark energy and dark matter, dominate over the other species, possibly suggesting Einstein's gravity extensions\footnote{For instance, the Planck satellite has not excluded Starobinsky's inflation, built up in terms of  a second order curvature correction to Hilbert-Einstein's action \cite{planck}.} \cite{referenza3,referenza3.1,referenza3bis}. Again, this poses new unexpected caveats jeopardizing Einstein's gravity, likely requiring new physics behind general relativity.
\noindent
These two energy domains are somewhat not fully-matched to each other and as a possible signature of such a problem could be attributed to the recently-stressed convoluted cosmic tensions \cite{referenza4,referenza4bis}. Certainly, the fact that measurements are poorly constrained due to the small number of data points severely limits our matching \cite{referenza5,referenza5bis,referenza5tris}. Though data catalogs are not large enough, constraining Einstein's gravity at different scales may culminate in groundbreaking discoveries and so several examples can be used to combine small and high red shift measures \cite{referenza6,referenza6a,referenza6b,referenza6c}. In this respect, as a possible example of technique that could be used for both low and high energy domains, we face the red and blue shift, based on local position invariance and precision test, along which red and blue shift measures between two identical clocks, regardless the clock structures, plays a crucial role in bounding how time intervals change in presence of gravity \cite{referenza7}.
\noindent
In this work, we consider the red and blue shift in three distinct contexts of low, intermediate and high gravity. Thereby, we employ a Schwarzschild - de Sitter solution, with its effective cosmological constant, and an axially symmetric Zipoy - Voorhees metric, where we constrain the prolate term, $\delta$, that enters the metric itself as signature of spherical symmetry departure. We get feasible spacetime free parameter constraints from suitable red and blue shift regimes and check the intervals of values these parameters may hold within astrophysical frameworks. To do so, we first assume Schwarzschild - de Sitter spacetime, with  $\Lambda$ bound got from Planck mission \cite{planck}, for neutron stars first (high gravity), then for white dwarfs (intermediate gravity) and finally for the Earth - Mars system (low gravity). As expected, the overall treatment provides red and blue shift incompatible values with those predicted by ongoing experiments in the solar system, but leads to slight acceptable bounds as neutron stars and white dwarfs are used as benchmarks. In analogy, we build up the same procedure for the Zipoy - Voorhees metric. Here, we check possible evidence for quadrupole corrections on the same astrophysical objects discussed for the Schwarzschild - de Sitter spacetime. Last but not least, we propose novel experimental setups to improve the quality of our outcomes. In addition, we propose the binary system Mars - Phobos, for working out satellite laser ranging to fix refined red and blue shift constraints. Summing up, we demonstrate the incompatibility between red and blue shifts at astrophysical level. However, we presume our technique to be used at a cosmological level to get experimental bounds  and then to compare the corresponding expectations with respect to those measures that cause tensions in cosmology.

\noindent
The paper is structured as follows. In Sec. \ref{sezione2}, we describe the main features of our method, i.e. showing how red and blue shift are characterized. In Sec. \ref{sezione3}, we relate our two spacetimes with the red and blue shifts. Thus, in Sec. \ref{sezione4}, we discuss our numerical results and new experimental configurations then, in Sec. \ref{sezione5}, we develop conclusions and perspectives of our work.


\section{The photon red and blue shift}\label{sezione2}

Red and blue shift represent a tool that measures the frequency modifications measured in the exchange of a photon between two observers. This technique does not involve the field equations but only the spacetime symmetries. In this section, we therefore analyze the general method used to determine the red and blue shift, assuming an emitted photon from a massive object, e.g. a planet, a star, a wormhole, and so forth and received by a distant observer. We first highlight the red and blue shift general treatment in a rotating frame. Then, we work out the same for a static configuration and we focus on how to use the underlying strategy for astrophysical objects.


\subsection{General treatment}
\label{subsec1}


Here we follow the mathematical procedure described in \cite{2.1,2.2} and start with the simplest metric describing a rotating  axially symmetric spacetime (in spherical coordinates)
\begin{equation}
\label{metric}
    ds^2=g_{tt}dt^2+2g_{t\varphi}dtd\varphi+g_{\varphi\varphi}d\varphi^2+g_{rr}dr^2+g_{\theta\theta}d\theta^2,
\end{equation}
where $g_{\mu\nu}=g_{\mu\nu}(r,\theta)$, in the gauge $g_{r\theta}=0$.
\noindent
Let us now indicate with $r_e$ and $r_d$ the photon's emitter and  photon's detector positions, respectively\footnote{From now on, to simplify the notation, we will denote these two positions with $r_p$, where the subscript $p$ can be $e$ or $d$ depending on whether it refers to the emitter or the detector respectively.} and so
$    u^{\mu}_{p}=(u^{t}_{p}, u^{r}_{p}, u^{\theta}_{p}, u^{\varphi}_{p})$
is the four-velocity of the photon's emitter (when $p=e$) or of the photon's detector (when $p=d$). Since the above relation refers to a  massive object, the following normalization condition holds
\begin{equation}
\label{norm1}
u^{\mu}_{p}u_{p,\mu}=-1\,,
\end{equation}
which explicitly reads
\begin{strip}
\begin{equation}
  \left.[g_{tt}(u^t)^2+g_{rr}(u^r)^2+g_{\varphi\varphi}(u^{\varphi})^2+g_{\theta\theta}(u^{\theta})^2+g_{t\varphi}u^t u^{\varphi}]\right|_{r=r_p}=-1\,.
\end{equation}
\end{strip}
\noindent
Similarly, we indicate with
$    k^{\mu}=(k^{t}, k^{r}, k^{\theta}, k^{\varphi})$
the four-velocity of the photon, albeit normalization condition is now
\begin{equation}
\label{norm2}
    k^{\mu}k_{\mu}=0\,,
\end{equation}
to explicitly give
\begin{eqnarray}
\nonumber
&&g_{tt}(k^t)^2+g_{rr}(k^r)^2+g_{\varphi\varphi}(k^{\varphi})^2+\\
&&+g_{\theta\theta}(k^{\theta})^2+g_{t\varphi}k^t k^{\varphi}=0.
\end{eqnarray}
Metric components are independent from the variables $t$ and $\varphi$, therefore there are two commuting Killing vector fields, respectively time-like and rotational ones as follow
\begin{eqnarray}
\label{killing1}
    &\xi^{\mu}=(1,0,0,0)\,, \\
    \label{killing2}
    &\psi^{\mu}=(0,0,0,1)\,.
\end{eqnarray}
These two Killing fields imply the existence of two conserved quantities for the massive particle
\begin{eqnarray}
    E&\doteq -g_{\mu\nu}\xi^{\mu}u^{\nu}=-g_{tt}u^t-g_{t\varphi}u^{\varphi}\,,\label{E} \\
     L&\doteq g_{\mu\nu}\psi^{\mu}u^{\nu}=g_{\varphi\varphi}u^{\varphi}+g_{t\varphi}u^t\,,\label{L}
\end{eqnarray}
that are the total energy,  $E$, and angular momentum, $L$.
\noindent
We now evaluate  $u^{\varphi}$ and $u^t$ in function of the energy, $E$, and angular momentum, $L$, from Eqs. (\ref{E}) and (\ref{L}), to give
\begin{eqnarray}
u^{\varphi}&=&-\frac{Eg_{t\varphi}+Lg_{tt}}{g^2_{t\varphi}-g_{\varphi\varphi}g_{tt}},\\
u^t&=&\frac{Eg_{\varphi\varphi}+Lg_{t\varphi}}{g^2_{t\varphi}-g_{\varphi\varphi}g_{tt}}\,,
\end{eqnarray}
and  plugging the above expressions  into Eq. (\ref{norm1}), we get:
\newpage
\begin{strip}
\begin{equation}
\label{norm3}
    \left.\left[g_{rr}(u^r)^2+g_{\theta\theta}(u^{\theta})^2+1-\frac{E^2g_{\varphi\varphi}+L^2g_{tt}+2ELg_{t\varphi}}{g^2_{t\varphi}-g_{\varphi\varphi}g_{tt}}\right]\right|_{r=r_p}=0.
\end{equation}
\end{strip}
\noindent
Even though the four-vector components for velocity and momentum do not vanish, rotating the polar coordinate system, the metric, Eq.  (\ref{metric}), does not change. Hence, this intrinsic symmetry implies that we can limit to the equatorial plane, where $\theta=\pi/2$, leading to  $u^{\theta}=k^{\theta}=0$.

Further, since we hereafter on circular orbits only, we even require  $u^r=0$, providing  Eq. (\ref{norm3}) becomes
\begin{equation}
    \label{potential}
    \left.\left[1-\frac{E^2g_{\varphi\varphi}+L^2g_{tt}+2ELg_{t\varphi}}{g^2_{t\varphi}-g_{\varphi\varphi}g_{tt}}\right]\right|_{r=r_p}=0,
\end{equation}
that reduces to
\begin{equation}
\label{Veff}
    V_{\rm{eff}}(r_p)=0.
\end{equation}
The former is the energy conservation law, clearly valid for circular orbits. In addition, these orbits require \cite{2.3,2.4}
\begin{equation}
    \label{Veff'}
    V_{\rm{eff}}'(r_p)=0\,,
\end{equation}
\begin{equation}
    \label{Veff''}
     V_{\rm{eff}}''(r_p)\geq 0\,,
\end{equation}
guaranteeing orbit stability \cite{2.4} and  the existence of the potential minimum\footnote{For the sake of completeness, the equality only holds  for spherical symmetry.}.
\noindent
\\Analogously, the two Killing fields, Eqs. (\ref{killing1}) and (\ref{killing2}), imply the existence of two conserved quantities for the photon, the total energy $E_{\gamma}$ and the angular momentum~$L_{\gamma}$
\begin{eqnarray}
\label{Egamma}
E_{\gamma}&\doteq& -g_{\mu\nu}\xi^{\mu}k^{\nu}=-g_{tt}k^t-g_{t\varphi}k^{\varphi}, \\
\label{Lgamma}
     L_{\gamma}&\doteq& g_{\mu\nu}\psi^{\mu}k^{\nu}=g_{\varphi\varphi}k^{\varphi}+g_{t\varphi}k^t.
\end{eqnarray}

\subsection{Evaluating the red and blue shift}

Now we have all the ingredients to determine the red and blue shift of the emitted photon. Thus, the photon frequency at given point $p$ is defined as \cite{2.5}
\begin{equation}
    \omega_p = -\left(k_{\mu}u^{\mu}\right)\vert_p.
\end{equation}
Since we consider timelike orbits that are both circular and equatorial, depending on whether we use Eqs. (\ref{E})-(\ref{L}) or Eqs. (\ref{Egamma})-(\ref{Lgamma}), we can rewrite $\omega_p$ in two ways, respectively:
\begin{eqnarray}
\omega_p&=&\left(Ek^t-Lk^{\varphi}\right)\vert_p,\\
\omega_p&=&\left(E_{\gamma}u^t-L_{\gamma}u^{\varphi}\right)\vert_p.
\end{eqnarray}
In particular, the frequency of the photon at the emission point is
\begin{eqnarray}
\nonumber
\omega_e &=& -\left(k_{\mu}u^{\mu}\right)\vert_e=\\
&=& \left(Ek^t-Lk^{\varphi}\right)\vert_e= \\
\nonumber
&=&\left(E_{\gamma}u^t-L_{\gamma}u^{\varphi}\right)\vert_e,
\end{eqnarray}
whereas the frequency of the photon at the detection point is
\begin{eqnarray}
\nonumber
\omega_d &=& -\left(k_{\mu}u^{\mu}\right)\vert_d=\\
&=& \left(Ek^t-Lk^{\varphi}\right)\vert_d= \\
\nonumber
&=&\left(E_{\gamma}u^t-L_{\gamma}u^{\varphi}\right)\vert_d.
\end{eqnarray}
Thus, we define the frequency shift associated with the emission and detection of photons as
\begin{eqnarray}
\label{a}
1+z=\frac{\omega_e}{\omega_d}&=&\frac{\left(E_{\gamma}u^t-L_{\gamma}u^{\varphi}\right)\vert_e}{\left(E_{\gamma}u^t-L_{\gamma}u^{\varphi}\right)\vert_d}=\nonumber\\
\label{fs}
    &=&\frac{\left(u^t-bu^{\varphi}\right)\vert_e}{\left(u^t-bu^{\varphi}\right)\vert_d},
\end{eqnarray}
where\footnote{Let us observe here that $b$ is the same both at the numerator and the denominator of (\ref{fs}), since $E_\gamma$ and $L_\gamma$ are determined by the same photon path.}
\begin{equation}\label{eq:b}
    b\equiv\frac{L_{\gamma}}{E_{\gamma}}\,.
\end{equation}
It will also be convenient to introduce the red shift $z_c$ corresponding to a photon emitted by a particle located at the center observed by a faraway detector, i.e. $b=0$:
\begin{equation}
\label{zc}
    z_c=\frac{u_e^t}{u_d^t}-1,
\end{equation}
since
astronomical data are generally collected in terms of the kinematic red shift, defined as
\begin{eqnarray}
\label{z_kin}
\nonumber
    z_{\rm{kin}}\doteq z-z_c&=&\frac{(u^t_e\,u^\varphi_d-u^t_d u^\varphi_e)b}{u^t_d(u_t-b u^\varphi)_d}\,=\\
    &=& \frac{(u^t_e\Omega_d-u^\varphi_e)b}{u^t_d(1-\Omega_db)}\,,\label{zkin}
\end{eqnarray}
where the \emph{angular velocity} of a detector located far away from the photons source
\begin{equation}
    \label{angularvel}
    \Omega_d\equiv\frac{u_d^{\varphi}}{u_d^t}
\end{equation}
 has been introduced as well.
\noindent
Of course, $b$ varies with the photon path. Then a value of $b$ as a function of the circular orbit of the emitting source (i.e., as a function of $r$) must be determined, in such a way that its absolute value represents the observed radial distance on either side of the observed center ($b=0$) by a faraway observer. The idea is that frequency shifts yielding maximum and minimum values correspond to photons emitted with initial velocities collinear to the source velocity \cite{NuSaSu2001}. This amounts to require $k^r=k^\theta=0$ at $p=e$, and therefore these photons paths, recalling (\ref{Egamma}), (\ref{Lgamma}), are such that
\begin{equation}
    -E_\gamma k^t_e+L_\gamma k^\varphi_e=(k_\mu\,k^\mu)_e=0,
\end{equation}
that gives, using (\ref{eq:b}), two possible solutions for
 the so called {\sl apparent impact parameter} $b$:
\begin{equation}
\label{bpm}
    b_{\pm}=-\frac{g_{t\varphi}\pm\sqrt{g_{t\varphi}^2-g_{\varphi\varphi}g_{tt}}}{g_{tt}},
\end{equation}
depending on whether the photon is emitted by a receding, $b_-$, or an approaching, $b_+$, object with respect to a distant observer. Hence, the $b_-$ and $b_+$ solutions are related to the red shift and the blue shift once we substitute them in Eq. (\ref{fs}), respectively
\begin{equation}
    \label{reds}
{\rm Red\,\, shift\,\,\,}    z_{\rm{red}}=\frac{u^t_e-b_{-}u^{\varphi}_e}{u^t_d\left(1-b_{-}\Omega_d\right)}-1,
\end{equation}
\begin{equation}
    \label{blues}
    {\rm Blue\,\, shift\,\,\,} z_{\rm{blue}}=\frac{u^t_e-b_{+}u^{\varphi}_e}{u^t_d\left(1-b_{+}\Omega_d\right)}-1.
\end{equation}
Finally, given $b_+$ and $b_-$, from Eq. (\ref{z_kin}), we get two possible $z_{kin}$ values, namely $z_1$ and $z_2$
\begin{equation}
    \label{z1}
    z_1=\frac{(u_e^t \Omega_d -u_e^{\varphi})b_{-}}{u_d^t(1-\Omega_d b_{-})},
\end{equation}
\begin{equation}
    \label{z2}
     z_2=\frac{(u_e^t \Omega_d -u_e^{\varphi})b_{+}}{u_d^t(1-\Omega_d b_{+})},
\end{equation}
that correspond to the cases in which the photon is emitted by a receding or an approaching source, respectively\footnote{It is remarkable to underline the relation between $z_{red}$, or $z_{blue}$, and $z_{kin}$. In particular, from Eq. (\ref{z_kin}), it reads
\begin{eqnarray}
 z_{red}&=&\left.(z_{kin}+z_c)\right|_{b=b_{-}},\\
 z_{blue}&=&\left.(z_{kin}+z_c)\right|_{b=b_{+}}.
\end{eqnarray}
}

\subsection{Non rotating spacetime}

As special case, we limit to  non rotating spacetimes, i.e. the ones for which $g_{t\varphi}=0$. This will be the case of Schwarzschild - de Sitter and Zipoy - Voorhees metrics that we analyze in the next sections. Thus, Eq.  (\ref{metric}) simply reduces to
\begin{equation}
\label{metric2}
    ds^2=g_{tt}dt^2+g_{\varphi\varphi}d\varphi^2+g_{rr}dr^2+g_{\theta\theta}d\theta^2,
\end{equation}
with gauge condition, $g_{r\theta}=0$. Clearly, all the previous equations before determined are accordingly simplified and so the conserved quantities associated to the massive particles (observes) now become
\begin{eqnarray}
E&=&-g_{tt}u^t,\\
L&=&g_{\varphi\varphi}u^{\varphi},
\end{eqnarray}
and so the velocities are
$u^t=-\frac{E}{g_{tt}}, u^{\varphi}=\frac{L}{g_{\varphi\varphi}}$, while the equation $
    V_{\rm{eff}}(r_p)=0$ for the effective potential becomes
\begin{equation}
\label{Veff_nr}
    \left.\left[1+\frac{E^2g_{\varphi\varphi}+L^2g_{tt}}{g_{\varphi\varphi}g_{tt}}\right]\right|_{r=r_p}=0.
\end{equation}
Similarly, the conserved quantities associated to the photon are
\begin{eqnarray}
E_{\gamma}&=&-g_{tt}k^t,\\
L_{\gamma}&=&g_{\varphi\varphi}k^{\varphi},
\end{eqnarray}
from which $
k^t=-\frac{E_{\gamma}}{g_{tt}}, k^{\varphi}=\frac{L_{\gamma}}{g_{\varphi\varphi}}$, so that the apparent impact parameter finally reads
\begin{equation}
\label{bnr}
    b_{\pm}=\mp\sqrt{-\frac{g_{\varphi\varphi}}{g_{tt}}}.
\end{equation}
The functional forms of $z_1$ and $z_2$ are identical to Eqs. (\ref{z1}) and (\ref{z2}) since assuming $g_{t\varphi}=0$ modifies only the apparent impact parameters rather than $z_{kin}$. We here observe that $b_{+}=- b_{-}$, implying  $z_1=-z_2$.
With the above recipe in our hand we are now in condition to handle spacetime symmetries to model astrophysical landscapes. We therefore report below the two metrics involved in our computation.

\section{Spacetime solutions}\label{sezione3}

Our purpose is to assess astrophysical frameworks by means of given spacetimes. Thereby, we first handle the simplest axisymmetric spacetime based on the Zipoy - Voorhees metric. We aim at modelling astrophysical objects, such as neutron stars and white dwarfs by means of such a metric. Afterwards, we switch to the spherical symmetry based on the Schwarzschild - de Sitter metric. In such a case, differently of the astrophysical case, we intend to work out cosmological scenarios and to compute red and blue shifts by fixing the cosmological constant from Planck's measurements \cite{planck}.
\noindent
Clearly, these two regimes, based on two different spacetime symmetries, are profoundly different from each other and, as above stated, we are therefore considering two distinct energy domains. The first is a regime of high gravity, since it deals with neutron stars and white dwarfs. The second is purely cosmological, involving infrared scales of energy. Below we first summarize each metric formalism and then we argue bounds over the free coefficients.

\subsection{The Zipoy - Voorhees metric}

The strategy of getting red and blue shift is here applied to the  \emph{Zipoy - Voorhees metric} \cite{2.6}. The metric, in  spherical coordinates, reads
\begin{equation}
    \label{ZVmetric}
    ds^2=-Fdt^2+\frac{1}{F}\left[Gdr^2+Hd\theta^2+(r^2-2kr)\sin^2\theta d\varphi^2\right],
\end{equation}
where
\begin{eqnarray}
\label{F}
F&=&\left(1-\frac{2k}{r}\right)^{\delta},\\
\label{G}
G&=&\left(\frac{r^2-2kr}{r^2-2kr+k^2\sin^2\theta}\right)^{\delta^2-1},\\
\label{H}
H&=&\frac{\left(r^2-2kr\right)^{\delta^2}}{\left(r^2-2kr+k^2\sin^2\theta\right)^{\delta^2-1}}.
\end{eqnarray}
Here, $\delta$ is a free parameter which can vary into three possible ranges
\begin{itemize}
    \item $\delta>1$:  tidal forces diverge at the singularity, particles are crushed;
    \item $\frac{1}{2}<\delta<1$: the singularity is mild, i.e. particles reach it with zero velocity;
    \item $\delta<\frac{1}{2}$: the singularity is repulsive, particles are ejected.
\end{itemize}
It is remarkable to notice the limiting case $\delta\rightarrow1$ provides the Schwarzschild metric, whereas $\delta\rightarrow{1/2}$ could show likely critical effects. For example, in Ref. \cite{naked} the authors worked out naked singularity configuration to get regions of repulsive gravity, using eigenvalue method \cite{autovalori1,autovalori2} and showing this interval as critical. However, we here focus on regular objects, such as NS, WD and/or solar system configurations, and so we do not expect any critical region over $\delta$ and/or red or blue shifts, as we effectively get later. Furthermore, $k=m/\delta$ is the ratio between the mass $m$ of the gravitational field and the $\delta$ parameter.
\noindent
As underlined in Sec. \ref{subsec1}, we are limiting to the equatorial plane, {\it i.e.},  $\theta=\pi/2$. The Zipoy - Voorhees metric describes a non rotating spacetime ($g_{t\varphi}=0$), thus we can consider Eq. (\ref{Veff_nr}) that reads
\begin{equation}
    \label{VeffZV}
    \left.1-\frac{E^2 \left(r^2-2 k r\right) \left(1-\frac{2 k}{r}\right)^{-\delta }-L^2 \left(1-\frac{2 k}{r}\right)^{\delta }}{r^2-2 k r}\right|_{r=r_p}=0
\end{equation}
Its derivative with respect to $r$ gives the condition for circular orbits, say Eq. (\ref{Veff'}):
\begin{strip}
\begin{equation}
\label{Veff'ZV}
\left.\frac{\left(1-\frac{2 k}{r}\right)^{-\delta } \left[2 \delta  k r E^2 (r-2 k)+2 L^2 (\delta  k+k-r) \left(1-\frac{2 k}{r}\right)^{2 \delta }\right]}{r^2 (r-2 k)^2}\right|_{r=r_p}=0.
\end{equation}
\end{strip}
\noindent
with $r_p=r_e,r_d$, as before.
Solving the system given by the two last relations, we obtain the total energy and the angular momentum
\begin{equation}
\label{E_ZV}
E=\left.\sqrt{\frac{\left(1-\frac{2 k}{r}\right)^{\delta } (\delta  k+k-r)}{2 \delta  k+k-r}}\right|_{r=r_p},
\end{equation}
\begin{equation}
\label{L_ZV}
L=\pm\left.\sqrt{\frac{\delta  k r (2 k-r) \left(1-\frac{2 k}{r}\right)^{-\delta }}{2 \delta  k+k-r}}\right|_{r=r_p}.
\end{equation}

\noindent Consequently, we immediately get
\begin{equation}
\label{utZV}
\left.u^t\right|_{r=r_p}=\left.-\sqrt{\frac{\left(1-\frac{2 k}{r}\right)^{-\delta } (\delta  k+k-r)}{2 \delta  k+k-r}}\right|_{r=r_p},
\end{equation}
\begin{equation}
\label{ufZV}
\left.u^{\varphi}\right|_{r=r_p}=\left.\pm \sqrt{\frac{\delta  k (2 k-r) \left(1-\frac{2 k}{r}\right)^{\delta }}{r (k-r)^2 (2 \delta  k+k-r)}}\right|_{r=r_p}.
\end{equation}
Furthermore, from Eq. (\ref{bnr}), we have
\begin{equation}
    b_{\pm}=\mp\frac{\sqrt{r^2-2kr}}{\left(1-\frac{2 k}{r}\right)^{\delta}}\,,
\end{equation}
Finally, substituting Eqs. (\ref{utZV}) - (\ref{ufZV}) evaluated in $r=r_d$ into Eq. (\ref{angularvel}), we get the angular velocity:
\begin{equation}
    \Omega_{d\pm}=\mp\sqrt{\frac{\delta  k (2 k-r_d) \left(1-\frac{2 k}{r_d}\right)^{2 \delta }}{r_d (k-r_d)^2 (\delta  k+k-r_d)}},
\end{equation}
where $\Omega_{d+}$ and $\Omega_{d-}$ are respectively referred to a co-rotating and to a counter-rotating photons source with respect to the angular velocity of the gravitational field source. In conclusion, substituting all these equations into Eqs. (\ref{z1}) - (\ref{z2}), we get the expressions for $z_1$ and $z_2$ for the Zipoy - Voorhees metric
\begin{strip}
\begin{eqnarray}
\label{z1_ZV}
\nonumber
z_{1\pm}&=&\pm\Biggl\{\left(1-\frac{2 k}{r_d}\right)^{\delta} \left(1-\frac{2 k}{r_e}\right)^{-\delta}\Biggl[-\sqrt{\frac{ \left(1-\frac{2 k}{r_e}\right)^{\delta} \left(1-\frac{2 k}{r_d}\right)^{2\delta}(r_d-2k)^2(k-r_e+k\delta)k\delta}{(r_d-k)^2(k-r_e+2k\delta)(r_d-k-k\delta)}}+\\
\nonumber
&&+\left(1-\frac{2 k}{r_d}\right)^{\delta}\sqrt{\frac{\left(1-\frac{2 k}{r_e}\right)^{\delta}(2k-r_e)^2 k\delta}{(k-r_e)^2(r_e-k-2k\delta)}}\Biggr]\Biggr\}\Biggl\{\sqrt{\frac{\left(1-\frac{2 k}{r_d}\right)^{\delta }(k-r_d+k\delta)}{k-r_d+2k\delta}}\Biggl[\left(1-\frac{2 k}{r_d}\right)^{\delta}+\\
&&\pm\sqrt{\frac{(r_d-2k)^2\left(1-\frac{2 k}{r_d}\right)^{2\delta}k\delta}{(r_d-k)^2(r_d-k-k\delta)}}\Biggr]\Biggr\}^{-1},
\end{eqnarray}
\end{strip}
\begin{strip}
\begin{eqnarray}
\label{z2_ZV}
\nonumber
z_{2\pm}&=&\pm\Biggl\{\left(1-\frac{2 k}{r_d}\right)^{\delta} \left(1-\frac{2 k}{r_e}\right)^{-\delta}\Biggl[\sqrt{\frac{ \left(1-\frac{2 k}{r_e}\right)^{\delta} \left(1-\frac{2 k}{r_d}\right)^{2\delta}(r_d-2k)^2(k-r_e+k\delta)k\delta}{(r_d-k)^2(k-r_e+2k\delta)(r_d-k-k\delta)}}+\\
\nonumber
&&-\left(1-\frac{2 k}{r_d}\right)^{\delta}\sqrt{\frac{\left(1-\frac{2 k}{r_e}\right)^{\delta}(2k-r_e)^2 k\delta}{(k-r_e)^2(r_e-k-2k\delta)}}\Biggr]\Biggr\}\Biggl\{\sqrt{\frac{\left(1-\frac{2 k}{r_d}\right)^{\delta }(k-r_d+k\delta)}{k-r_d+2k\delta}}\Biggl[\left(1-\frac{2 k}{r_d}\right)^{\delta}+\\
&&\mp\sqrt{\frac{(r_d-2k)^2\left(1-\frac{2 k}{r_d}\right)^{2\delta}k\delta}{(r_d-k)^2(r_d-k-k\delta)}}\Biggr]\Biggr\}^{-1},
\end{eqnarray}
\end{strip}
\noindent
where the subscript $\pm$ is again referred to a co-rotating and counter-rotating source with respect to the angular velocity of the gravitational field source.
\noindent
Let us observe that $z_1=-z_2$, in both the co-rotating and counter-rotating cases, regardless of the mass that generates the gravitational field.
Above we put forward that the $z_1$ and $z_2$ variations can be expressed in terms of $r_d$ for both a rotating and counter-rotating configurations. This would help to argue the intervals of validity for the Zipoy - Voorhees free parameters when this metric is applied to astrophysical situations. We describe this approach in detail below.

\subsection{Gravitational field sources for the Zipoy - Voorhees metric}

\begin{figure}
    \begin{center}
    \includegraphics[width=1.07\columnwidth]{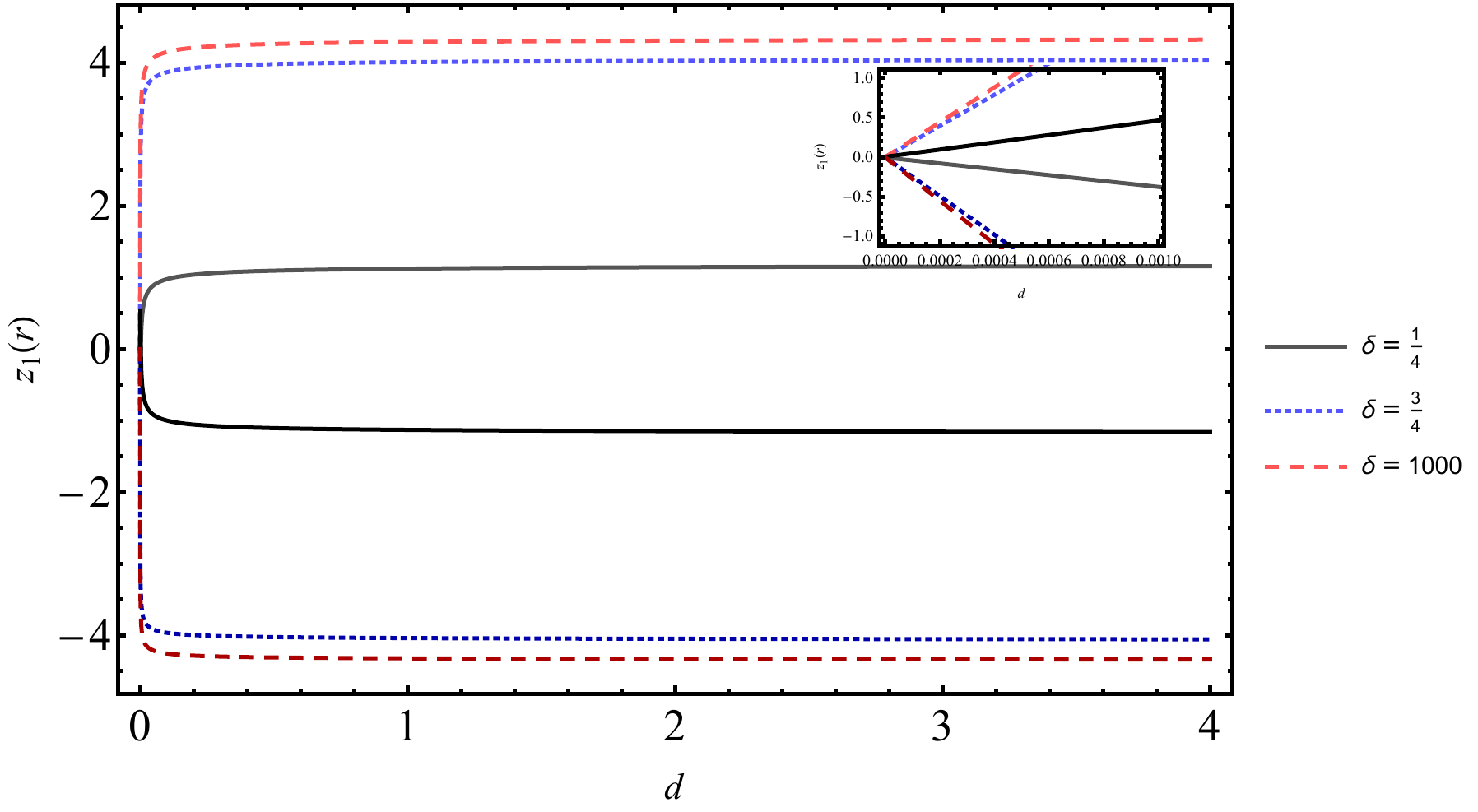}
    \includegraphics[width=1.07\columnwidth]{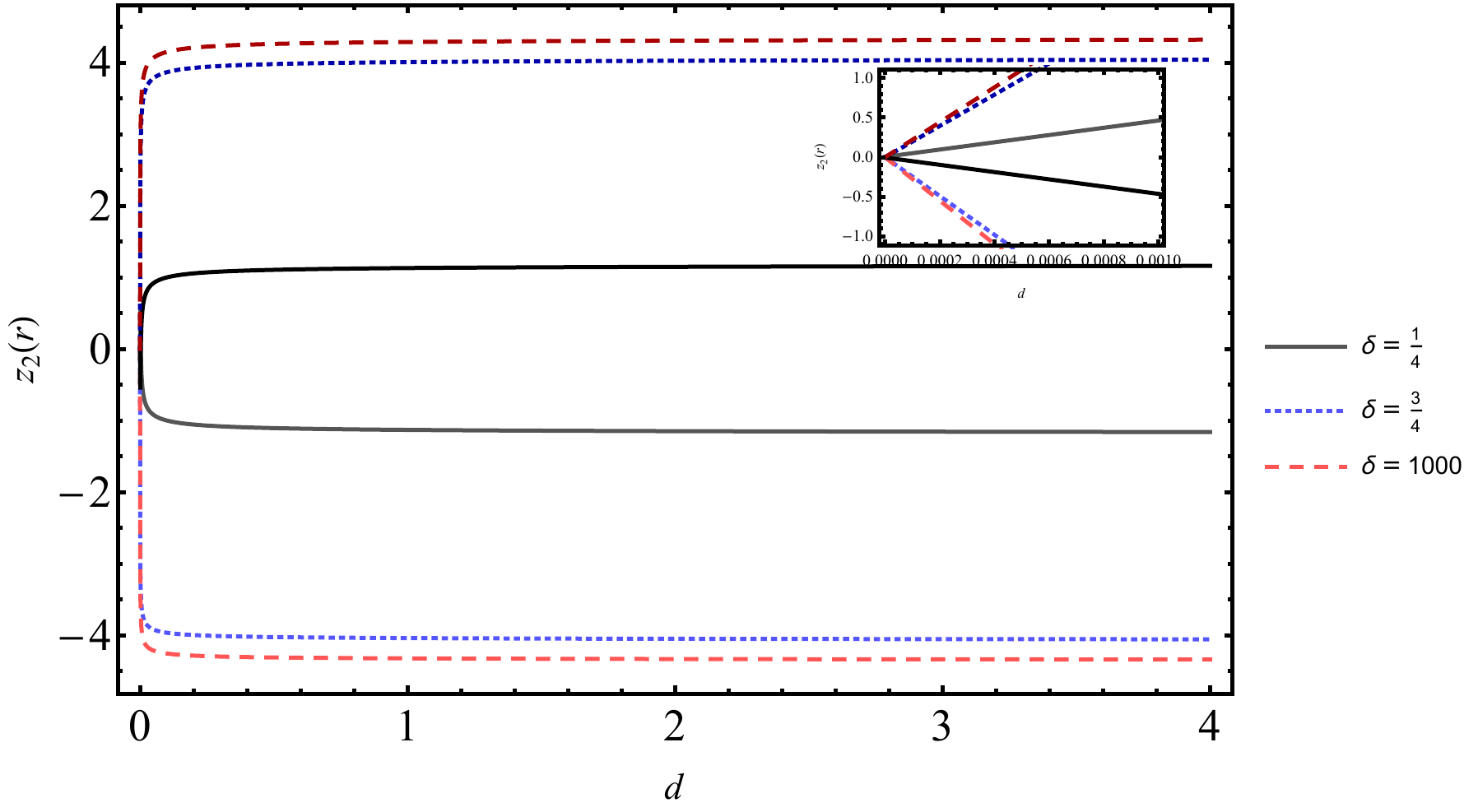}
    \caption{    \emph{$z_1(r)$ and $z_2(r)$ as function of $d=r_d-r_e$ within the Zipoy - Voorhees spacetime. The gravitational field source is a neutron star of mass $M=1.07M_\odot$, with $M_\odot=1.47$ km the solar mass, and radius $R=13.61$ km, in the maximally - rotating configuration \cite{2.7}. Here,  $d\in[0;4\cdot 10^5]$ km, whereas  $z_1$ and $z_2$ are in power of $10^{-1}$. In the small zoom, we report  $z_1$ and $z_2$ up to $d=40$ km.}}
    \label{fig:NSZV}
\end{center}
\end{figure}

\begin{figure}
    \begin{center}
    \includegraphics[width=0.9\columnwidth]{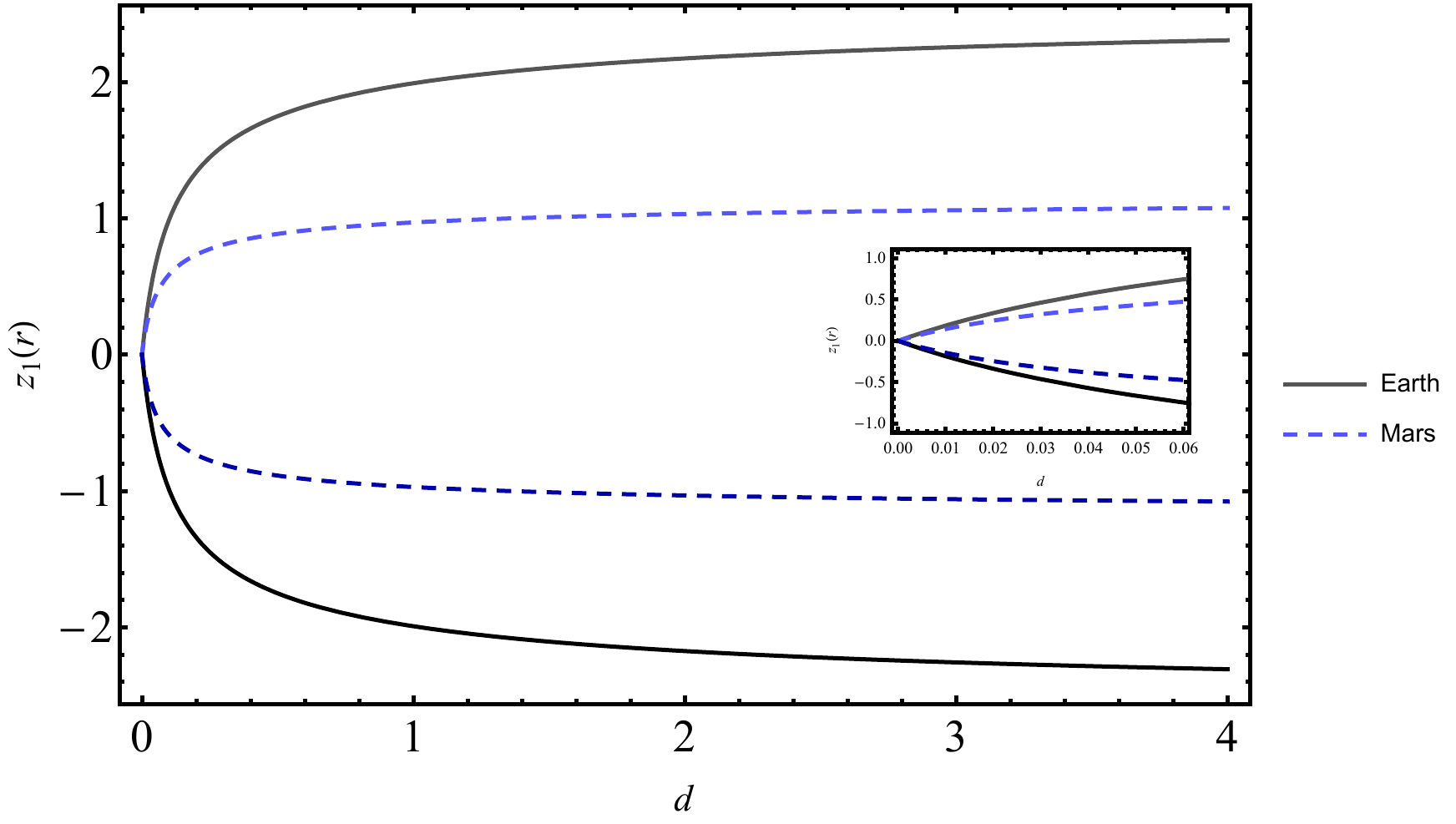}
    \includegraphics[width=0.9\columnwidth]{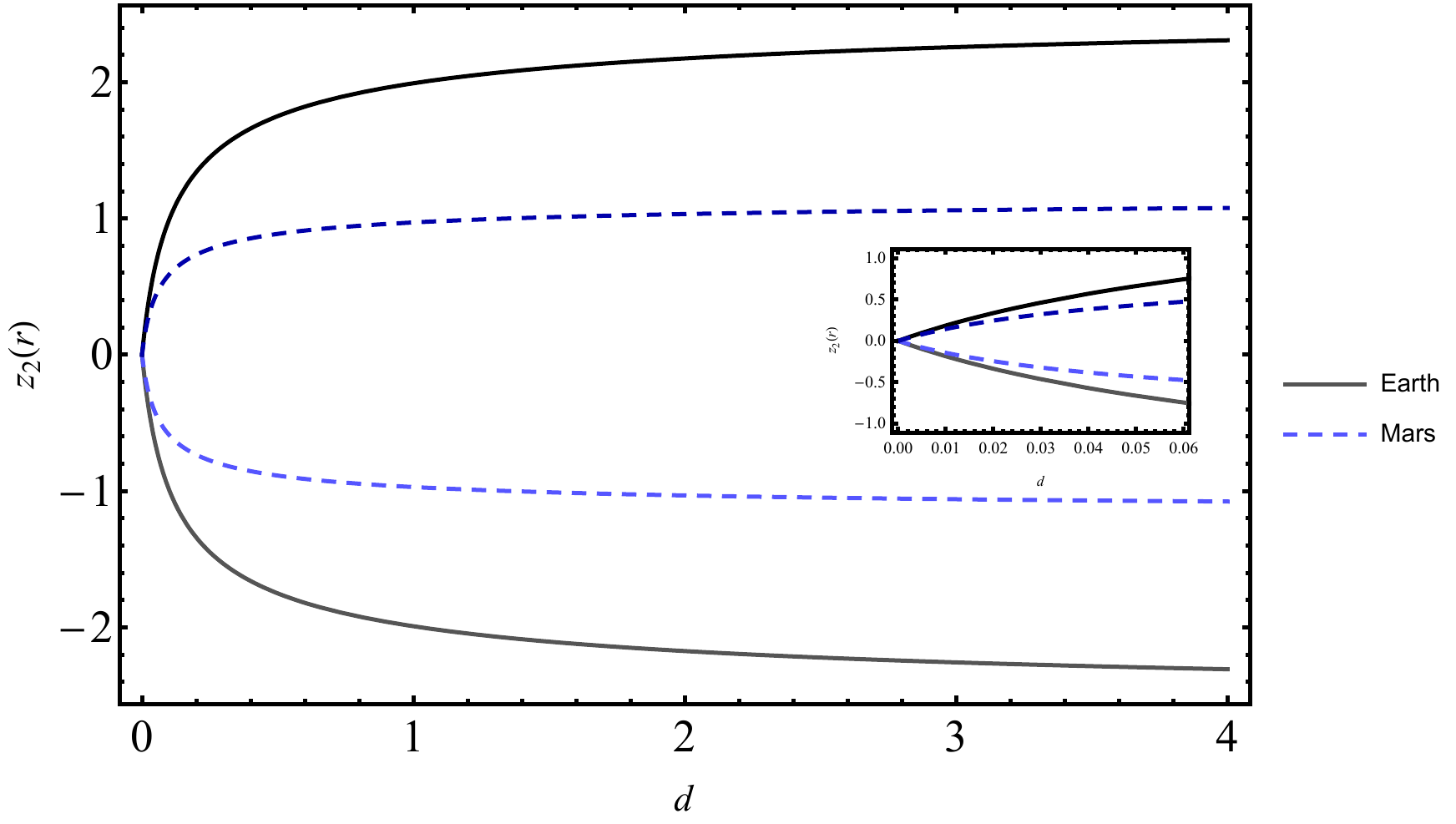}
    \caption{\emph{$z_1(r)$ and $z_2(r)$ as function of $d=r_d-r_e$ within the Zipoy - Voorhees spacetime. We analyzed the two cases, i.e. emitter on Mars and Earth respectively, with $d\in[0;4\cdot 10^5]$ km, whereas  $z_1$ and $z_2$ are in power of $10^{-5}$. In the small zoom, we report  $z_1$ and $z_2$ up to $d=6\cdot 10^3$ km. This choice enables to get feasible intervals for Phobos, where $d\simeq 5986.5$ km, with Mars as source, and intervals for LARES2 \cite{lares}, where  $d\simeq 5899$ km, with Earth as source.}}
       \label{fig:EarthMarsZV}
    \end{center}
\end{figure}

We analyze the variation of $z_1$ and $z_2$ as function of the position of the detector $r_d$, in the co-rotating and in the counter-rotating configurations. Our analysis is based on different gravitational field sources
\begin{itemize}
    \item a neutron star in the maximally - rotating configuration \cite{2.7}, corresponding to a high gravity regime,
    \item a white dwarf in the maximally - rotating configuration \cite{2.7}, correspoding to an intemediate gravity regime,
    \item Earth and Mars for the Solar System, corresponding to a low gravity regime.
\end{itemize}
\noindent
We report the plots \ref{fig:NSZV}, \ref{fig:WDZV} and \ref{fig:EarthMarsZV} in which we infer the availability intervals for each term.
\\For the neutron star, the variation of $z_1$ and $z_2$ as function of $r_d$ depend stronger on $\delta$. For this reason, we choose three value of $\delta$, one for each range, Fig. \ref{fig:NSZV}
\begin{itemize}
    \item[{\bf 1.}] $\delta=1000$, {\it i.e.}, where we take an arbitrary large value to address the condition $\delta\gg 1$  ,
    \item[{\bf 2.}] $\delta=\frac{3}{4}$, as arbitrary close value to $\delta=1$, obtained as mean value of the interval $\frac{1}{2}<\delta<1$,
    \item[{\bf 3.}]  $\delta=\frac{1}{4}$, as arbitrary close value to $\delta =0$, obtained as mean value of the interval $0<\delta<\frac{1}{2}$.
\end{itemize}
However, for WDs, see Fig. \ref{fig:WDZV}, the increase or decrease of $\delta$ do not seem to modify the overall evolution. The same happens for the  binary configuration constituted by the Earth and Mars: one can notice from Fig. \ref{fig:EarthMarsZV} that they very weakly depend upon $\delta$ variation. The above configuration is built up assuming the Earth and Mars as distinct gravitational sources as separate cases. For the sake of clearness, the $\delta$ variation is not so evident from our plots since those variations are extremely small and not particularly visible. The corresponding values have been evaluated for WDs, Earth and Mars, noticing a slight difference that permits one to fix $\delta$ to portray the examples we showed in the aforementioned figures.
\noindent
Even though not so evident from our plots, the above occurrence for which $\delta$ is as larger as one approaches higher gravity regimes turns out to be clear even from a theoretical viewpoint. As one approaches regimes of low gravity any quadrupole deviation is negligibly small and so one can approximate with a spherical symmetry those configurations, without losing generality. Furthermore, in the low and intermediate gravity regimes the symmetries $z_{1+}=z_{2-}$ and $z_{1-}=z_{2+}$ emerge, together with  $z_{1\pm} =-z_{2\pm}$, being valid for any gravitational sources.
\noindent
The pending caveat to check would be represented by orbit stability, i.e. Eq. (\ref{Veff''}). For the Zipoy - Voorhees metric, the second derivative with respect to $r$ of Eq. (\ref{VeffZV}) is always zero: since this is a quasi-spherical spacetime, we can assert that all orbits are stable.
\\Finally, let us observe that, for $\delta=1$ and, consequently, $k=m$, all these equations reduce to those obtained in the Schwarzschild metric (see Appendix \ref{appendix}). It is now remarkable to stress that for $\delta=1/2$, in all the analyzed gravity regimes, we do not obtain critical values of z, as expected.

\subsection{The  Schwarzschild - de Sitter metric}

In this subsection, we apply the method described above to the Schwarzschild - de Sitter metric, corresponding to a spherical symmetric spacetime with an effective cosmological constant, $\Lambda$ \cite{SdSpaper1,SdSpaper2,SdSpaper3}. For this fundamental property, the metric can be used for cosmological applications, to infer bounds on red and blue shifts, fixing $\Lambda$. In spherical coordinates, we have
\newpage
\begin{strip}
\begin{equation}
    \label{metric_SdS}
    ds^2=-\left(1-\frac{2m}{r}+\frac{\Lambda r^2}{3}\right)dt^2+\frac{1}{\left(1-\frac{2m}{r}+\frac{\Lambda r^2}{3}\right)}dr^2+r^2d\theta^2+r^2\sin^2\theta d\varphi^2.
\end{equation}
\end{strip}
\noindent
We again stress, as we did in Sec. \ref{subsec1}, we study the phenomenon in the equatorial plane, namely $\theta=\pi/2$. Even this metric is clearly non-rotating and so Eq. (\ref{Veff_nr}) becomes
\begin{equation}
    \label{VeffSdS}
    \left.\left[1-\frac{E^2r^2-L^2\left(1-\frac{2m}{r}+\frac{\Lambda r^2}{3}\right)}{r^2\left(1-\frac{2m}{r}+\frac{\Lambda r^2}{3}\right)}\right]\right|_{r=r_p}=0\,,
\end{equation}
and its derivative with respect to $r$
\begin{equation}
    \label{Veff'SdS}
    \left.\left[\frac{6(3m+\Lambda r^3)E^2}{(\Lambda r^3+3r-6m)}-\frac{2L^2}{r^3}\right]\right|_{r=r_p}=0.
\end{equation}
with $r_p=\left\{r_e,r_d\right\}$, as before.
Solving the system given by the former two equations, we obtain
\begin{eqnarray}
\label{E_SdS}
E^2&=&\left.\frac{(\Lambda r^3+3r-6m)^2}{9r(r-3m)}\right|_{r=r_p},\\
\label{L_SdS}
L^2&=&\left.\frac{r^2(3m+\Lambda r^3)}{3(r-3m)}\right|_{r=r_p},
\end{eqnarray}
from which we get the total energy and the angular momentum:
\begin{eqnarray}
E&=&\left.\frac{\Lambda r^3+3r-6m}{\sqrt{9r(r-3m)}}\right|_{r=r_p},\\
L&=&\left.\pm r\sqrt{\frac{3m+\Lambda r^3}{3(r-3m)}}\right|_{r=r_p}.
\end{eqnarray}
Thus, we again find
\begin{equation}
    \label{ut_SdS}
    \left.u^t\right|_{r=r_p}=\left.\sqrt{\frac{r}{r-3m}}\right|_{r=r_p},
\end{equation}
\begin{equation}
    \label{uf_SdS}
    \left.u^{\varphi}\right|_{r=r_p}=\left.\pm\sqrt{\frac{3m+\Lambda r^3}{3r^2(r-3m)}}\right|_{r=r_p}.
\end{equation}
Furthermore, we have
\begin{equation}
    b_{\pm}=\mp\sqrt{\frac{r^2}{\left(1-\frac{2m}{r}+\frac{\Lambda r^2}{3}\right)}},
\end{equation}

\begin{equation}
    \Omega_{d\pm}=\pm\sqrt{\frac{3m+\Lambda r_d^3}{3r_d^3}},
\end{equation}
where $\Omega_{d+}$ and $\Omega_{d-}$ are respectively referred to a co-rotating and to a counter-rotating photons source with respect to the angular velocity of the gravitational field source, as before.
\noindent
Hence, plugging all these relations within Eqs. (\ref{z1})-(\ref{z2}), we get the expressions for $z_1$ and $z_2$ for the Schwarzschild - de Sitter metric
\begin{strip}
\begin{eqnarray}
\label{z1_SdS}
z_{1\pm}&=&\pm\frac{\sqrt{\frac{r_e}{r_e-3m}}\left(\sqrt{\frac{3m+\Lambda r_d^3}{3r_d-6m+\Lambda r_d^3}}-\sqrt{\frac{3m+\Lambda r_e^3}{3r_e-6m+\Lambda  r_e^3}}\right)}{\sqrt{\frac{r_d}{r_d-3m}}\left(1\mp\sqrt{\frac{3m+\Lambda r_d^3}{3r_d-6m+\Lambda r_d^3}}\right)},\\
\,\nonumber\\
z_{2\pm}&=&\mp\frac{\sqrt{\frac{r_e}{r_e-3m}}\left(\sqrt{\frac{3m+\Lambda r_d^3}{3r_d-6m+\Lambda r_d^3}}-\sqrt{\frac{3m+\Lambda r_e^3}{3r_e-6m+\Lambda r_e^3}}\right)}{\sqrt{\frac{r_d}{r_d-3m}}\left(1\pm\sqrt{\frac{3m+\Lambda r_d^3}{3r_d-6m+\Lambda r_d^3}}\right)},\label{z2_SdS}
\end{eqnarray}
\end{strip}
where the subscript $\pm$ is again referred to a co-rotating and counter-rotating source with respect to the angular velocity of the gravitational field source. Let us observe that $z_1=-z_2$, in both the co-rotating and counter-rotating cases, regardless of the mass that generates the gravitational field.

\subsection{Gravitational field sources for the Schwarzschild - de Sitter metric}

\begin{figure}
    \includegraphics[width=0.9\columnwidth]{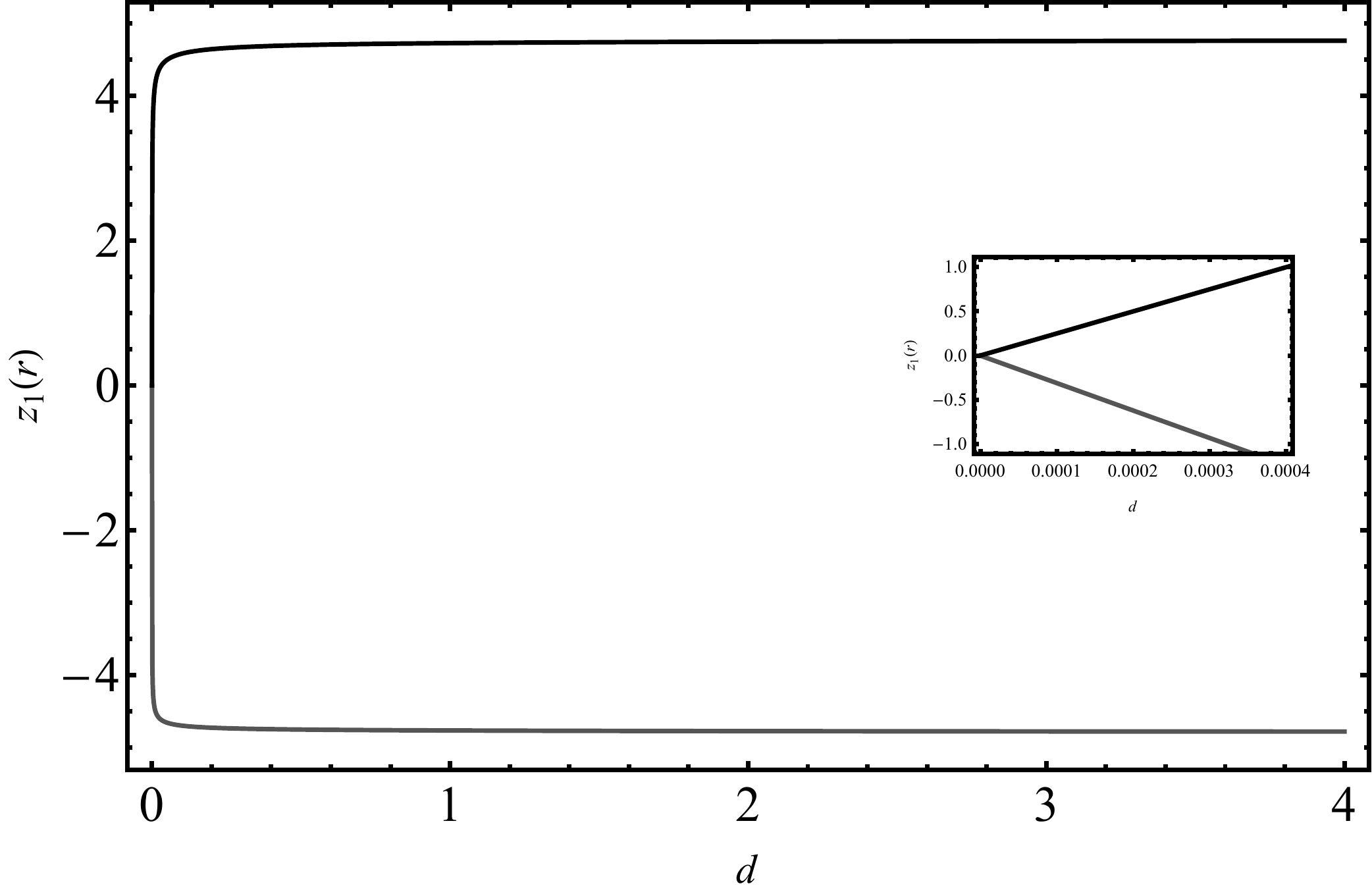}
    \includegraphics[width=0.9\columnwidth]{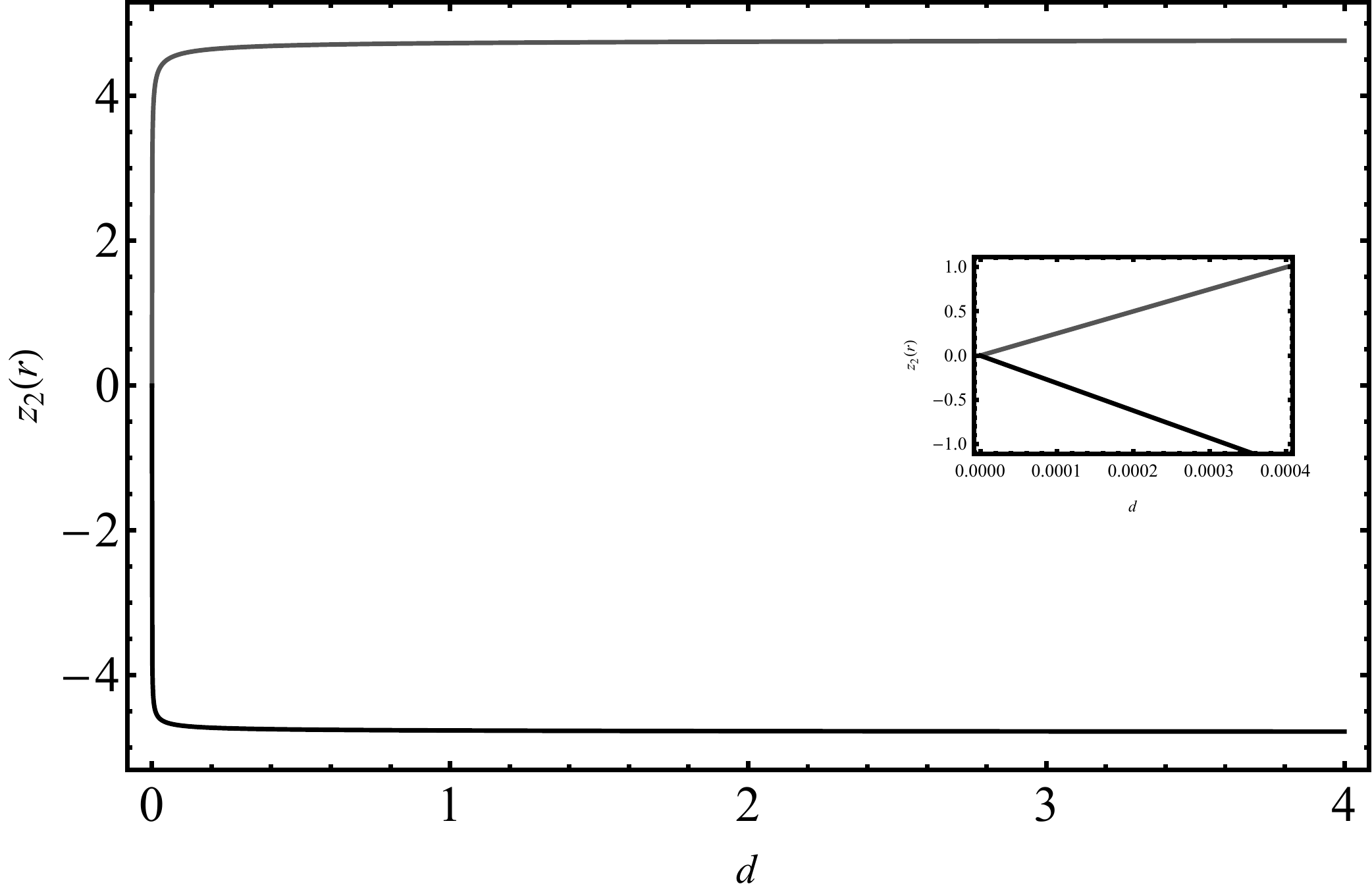}
    \caption{    \emph{$z_1(r)$ and $z_2(r)$ as function of $d=r_d-r_e$ within the Schwarzschild - de Sitter spacetime. The gravitational field source is a neutron star of mass $M=1.07M_\odot$, with $M_\odot=1.47$ km the solar mass, and radius $R=13.61$ km, in the maximally - rotating configuration \cite{2.7}. Here,  $d\in[0;4\cdot 10^5]$ km, whereas  $z_1$ and $z_2$ are in power of $10^{-1}$. In the small zoom, we report  $z_1$ and $z_2$ up to $d=40$ km.}}
    \label{fig:NSSDS}
\end{figure}

\begin{figure}
    \begin{center}
    \includegraphics[width=0.9\columnwidth]{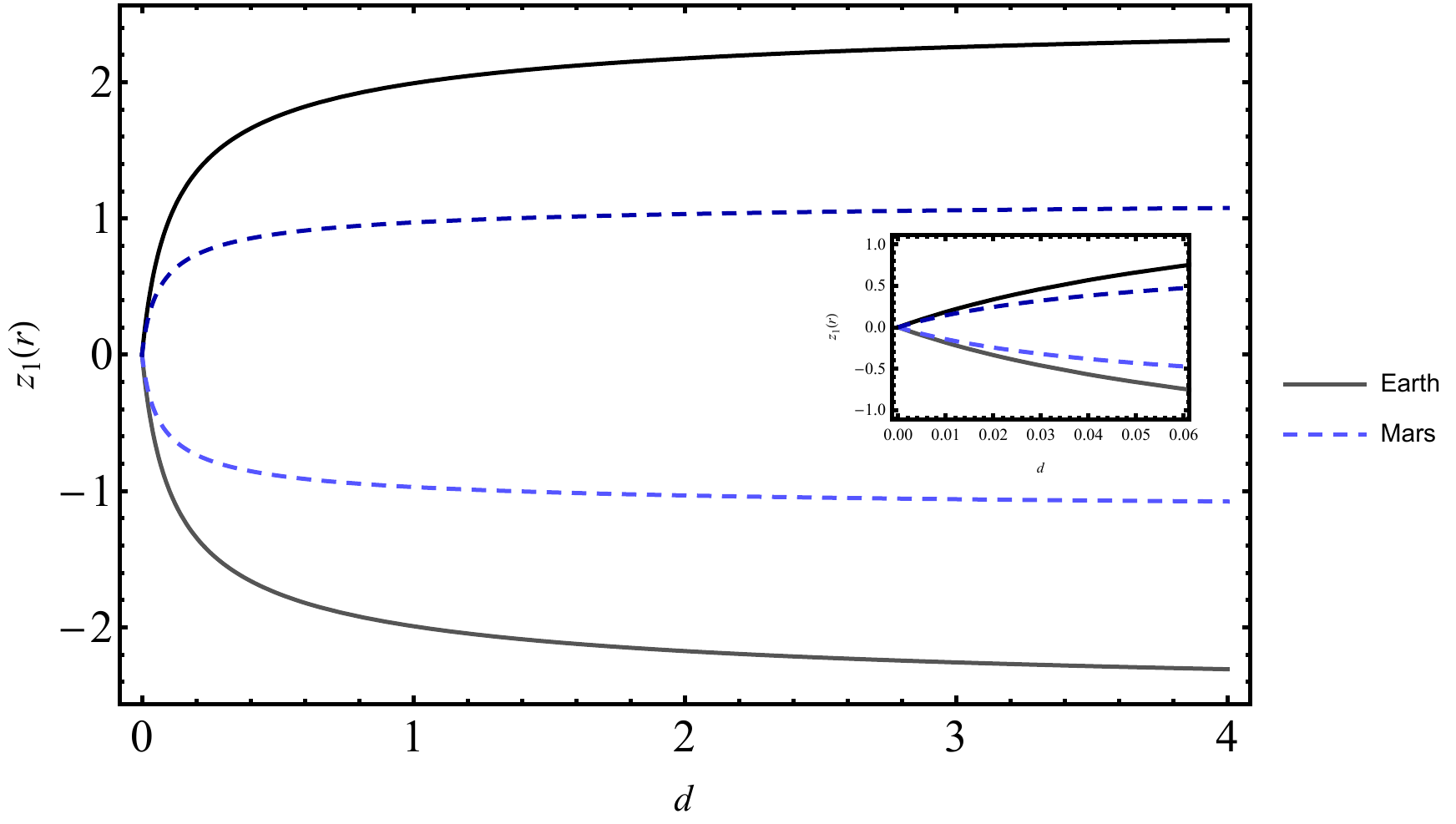}
    \includegraphics[width=0.9\columnwidth]{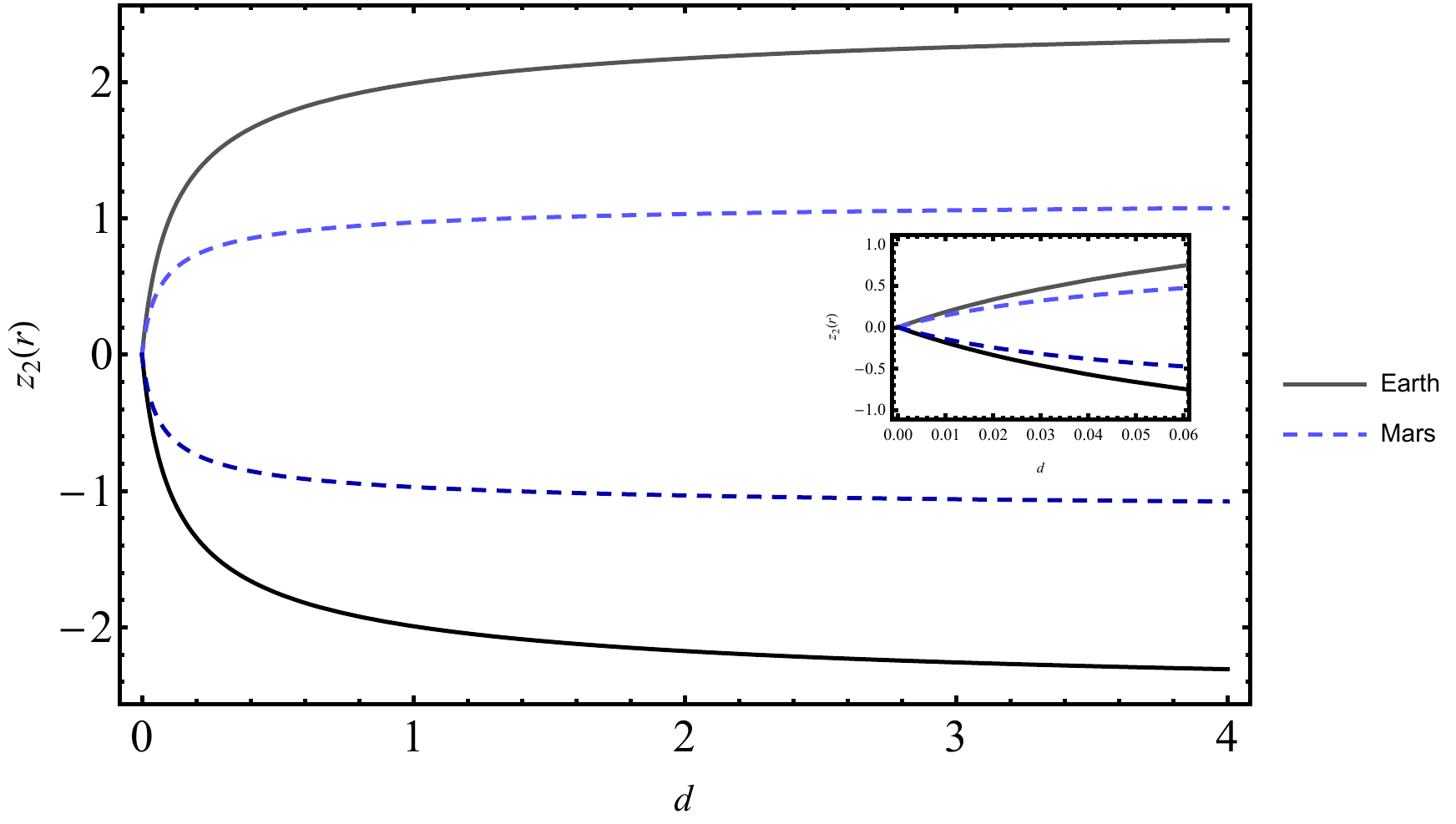}
    \caption{    \emph{$z_1(r)$ and $z_2(r)$ as function of $d=r_d-r_e$ within the Schwarzschild - de Sitter spacetime. We analyzed the two cases, i.e. emitter on Mars and Earth respectively, with $d\in[0;4\cdot 10^5]$ km, whereas  $z_1$ and $z_2$ are in power of $10^{-5}$. In the small zoom, we report  $z_1$ and $z_2$ up to $d=6\cdot 10^3$ km. This choice enables to get feasible intervals for Phobos, where $d\simeq 5986.5$ km, with Mars as source, and intervals for LARES2 \cite{lares}, where  $d\simeq 5899$ km, with Earth as source.}}\label{fig:EarthMarsSDS}
    \end{center}
\end{figure}

In analogy to our previous treatment, for the Schwarzschild - de Sitter metric we analyze the variation of $z_1$ and $z_2$ as function of the position of the detector $r_d$, in the co-rotating and in the counter-rotating cases. As before, our analysis is based on different gravitational field sources:
\begin{itemize}
    \item a neutron star in the maximally - rotating configuration \cite{2.7}, see Fig. \ref{fig:NSSDS}. Here the employed field is strong,
    \item a white dwarf in the maximally - rotating configuration \cite{2.7}, see Fig. \ref{fig:WDSDS}. Here we consider an intermediate field,
    \item Earth and Mars for the Solar System, see  Fig. \ref{fig:EarthMarsSDS}.
\end{itemize}
Furthermore, we consider $\Lambda$ as the cosmological constant, whose value is $\Lambda=1.1056\times 10^{-52}\,m^{-2}$ \cite{planck}.
\noindent
As for the Zipoy - Voorhees metric, in the low and intermediate gravity regimes we again find $z_{1+}=z_{2-}$ and $z_{1-}=z_{2+}$, in addition to $z_{1\pm} =-z_{2\pm}$.
\noindent
The last thing to check is the orbits stability, i.e. Eq. (\ref{Veff''}). For the Schwarzschild - de Sitter metric, the second derivative of Eq. (\ref{VeffSdS}) is
\begin{equation}
\label{V''effSdS}
    \left.\left[\frac{8\Lambda r^4+6mr(1-5\Lambda r^2)-36m^2}{r^2(r-3m)(\Lambda r^3+3r-6m)}  \right]\right|_{r=r_p}\geq 0.
\end{equation}
Thus, we study Eq. (\ref{V''effSdS}) for every gravitational field source.
\\By considering the photons emitter placed on the neutron star, i.e. $r_e=R$, with $R$ the neutron star radius, we get that the emitter orbit is stable:
\begin{equation}
    V''_{\rm{eff}}(r_e)=7.61641\times 10^{-4}>0,
\end{equation}
while the detector orbit is stable for:
\begin{equation}
    r_d\geq 9.4374\;\rm{km}.
\end{equation}
Analogously, by considering the photons emitter placed on the white dwarf, i.e. $r_e=R$, with $R$ the white dwarf radius, we get that the emitter orbit is stable:
\begin{equation}
    V''_{\rm{eff}}(r_e)=8.62859\times 10^{-14}>0,
\end{equation}
while the detector orbit is stable for:
\begin{equation}
    r_d\geq 1.5876\;\rm{km}.
\end{equation}
Finally, by considering the photons emitter placed on the planet, i.e. $r_e=r_{\rm{Earth}}$ and $r_e=r_{\rm{Mars}}$, with $r_{\rm{Earth}}$ and $r_{\rm{Mars}}$ the Earth and Mars radius respectively, we get that the emitters orbits are stable
\begin{eqnarray}
V''_{\rm{eff}}(r_e)&=&3.42303\times 10^{-17}>0\;\;\mbox{for  Earth},\\
V''_{\rm{eff}}(r_e)&=&2.43224\times 10^{-17}>0\;\;\mbox{for Mars},
\end{eqnarray}
while the detectors orbits are stable for
\begin{eqnarray}
r_d&\geq&2.65555\times 10^{-5}\;\rm{km}\;\;\mbox{for  Earth},\\
r_d&\geq&2.84142\times 10^{-6}\;\rm{km}\;\;\mbox{for Mars}.
\end{eqnarray}
As before, let us observe that, for $\Lambda\rightarrow0$, all these equations reduce to those obtained in the Schwarzschild metric. For the sake of completeness, we briefly report the details in Appendix \ref{appendix}.

\section{Theoretical discussion}
\label{sezione4}

In this section, we describe our findings confronting our predictions with current experimental bounds, got from experiments. Further, we propose how to build up plausible experiments and develop technical configurations to check the validity of our methods. We first discuss our outcomes in the two perspectives that we described above, {\it i.e.}, Zipoy - Voorhees and Schwarzschild - de Sitter metrics. Then, we highlight the basic demands of likely experimental features to check the goodness of our limits.

\subsection{High and intermediate gravity regimes}

The increase or decrease of our red and blue shift bounds depend upon the choice of our free parameters. The possible underlying configuration is crucial in  understanding how to single out the most feasible interval of red shifts or blue shifts. We then split the involved two symmetries below, commenting separately our findings and comparing our bounds with previous expectations got from the literature.
\begin{table}[htp]
\begin{tabular}{ccc}
\hline
\hline
         & Numerical values got during computation &    \\
\hline\hline
         & M  &  R  \\
  &  ($M_\odot$)  & ($10^3 m$) \\
  \hline
  WD       & $0.18$ &   $18304.5$  \\
  \hline
  NS  & $1.07$  &  $13.61$  \\
\hline
\hline
\end{tabular}
\vspace{0.3cm}
\label{table1}
\caption{\emph{Table of astrophysical values adopted during our computation for high and intermediate gravity regimes. We only consider the maximally rotating configurations for NS and WD, where gravitational effects are stringent.}}
\end{table}

\subsubsection{NS case}

In the case of NS, we compute our expectations over $z_1(r)$ and $z_2(r)$ in the maximally - rotating configuration, with the ranges of masses and radii respectively given by $M\in[1.07;1.47]M_\odot$ and $r\simeq 13.61$. The values got by $z_1$ and $z_2$ reach a plateau as $d\gtrsim0$,  i.e., as $d$ becomes larger than zero. This indicates the strong gravity regime of NS, as expected, and happens for both the setups of Schwarzschild - de Sitter and Zipoy - Voorhees spacetimes, when the former is computed for very large $\delta$ values. Particularly, the Zipoy - Voorhees metric seems to match the Schwarzschild - de Sitter solution as the quadrupole increases, in agreement with the fact that NS are described as rotating objects. Further, this indicates the Schwarzshild - de Sitter spacetime is a suitable approximation for determining the NS red and blue shifts, although the metric itself does not describe a rotating object. The very impressive fact is that one underlines very small changes within the interval $\delta \in[0.75;10^3]$. This suggests a limiting regime between the above interval, being compatible with the theoretical bounds over  $\delta$ that exclude repulsive effects of gravity.

\subsubsection{WD case}

\begin{figure}
    \includegraphics[width=0.9\columnwidth]{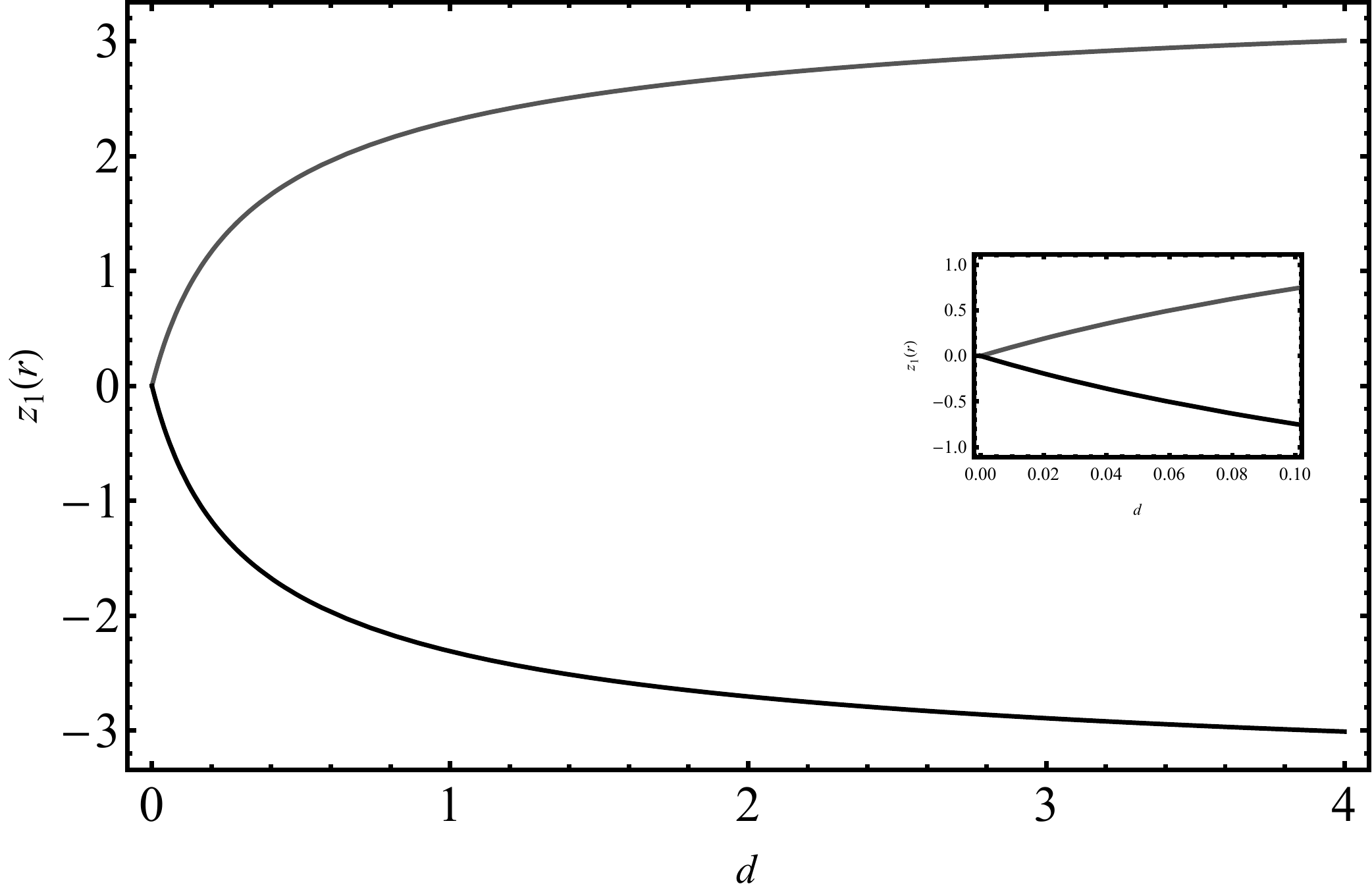}
    \includegraphics[width=0.9\columnwidth]{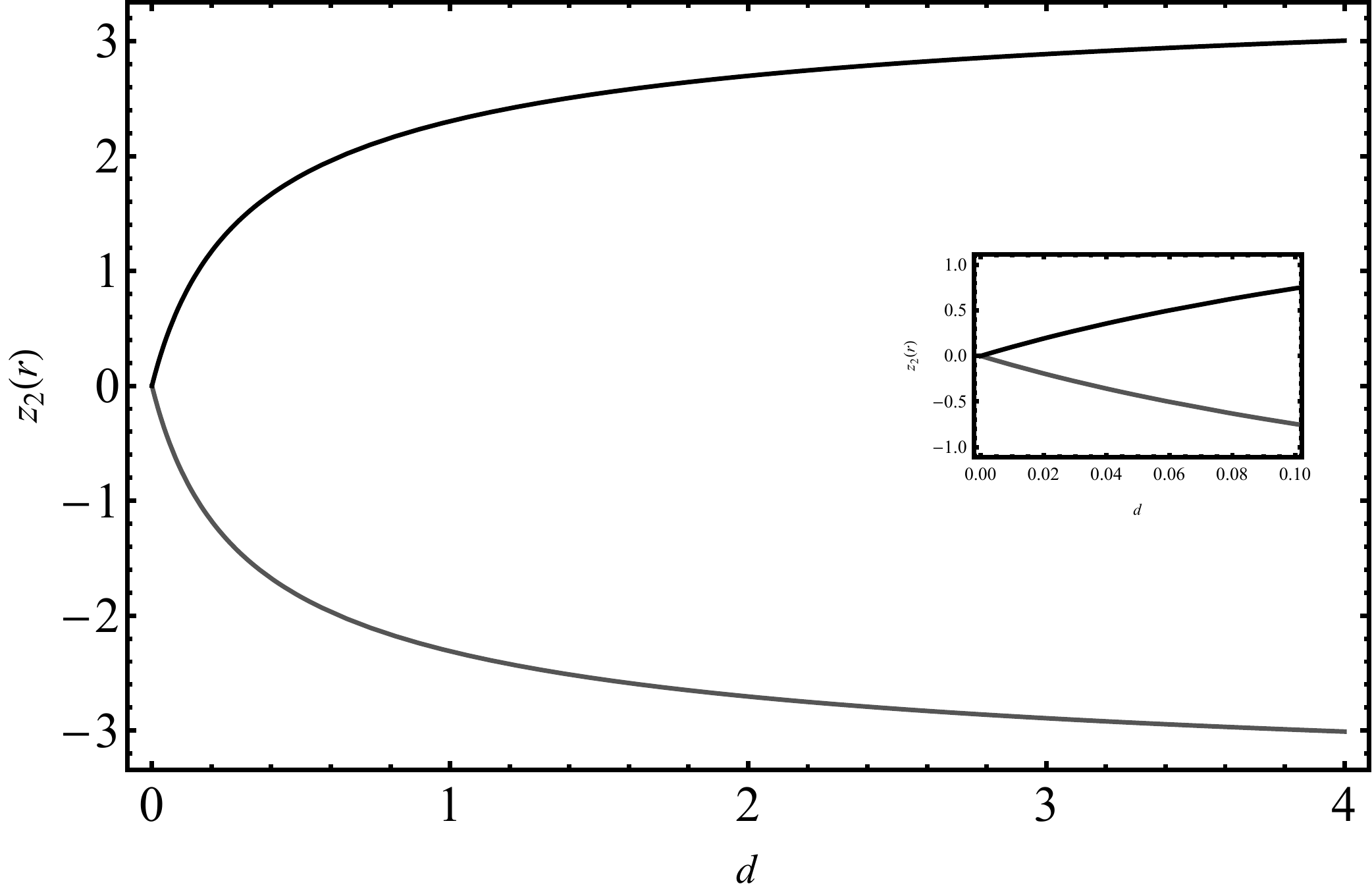}
    \caption{  \emph{$z_1(r)$ and $z_2(r)$ as function of $d=r_d-r_e$ within the Zipoy - Voorhees spacetime. The gravitational field source is a white dwarf of mass $M=0.18M_\odot$, with $M_\odot=1.47$ km the solar mass, and radius $R=18304.5$ km, in the maximally - rotating configuration \cite{2.7}. Here,  $d\in[0;4\cdot 10^5]$ km, whereas  $z_1$ and $z_2$ are in power of $10^{-3}$. In the small zoom, we report  $z_1$ and $z_2$ up to $d=10^4$ km.}}      \label{fig:WDZV}
\end{figure}

\begin{figure}
    \begin{center}
    \includegraphics[width=0.9\columnwidth]{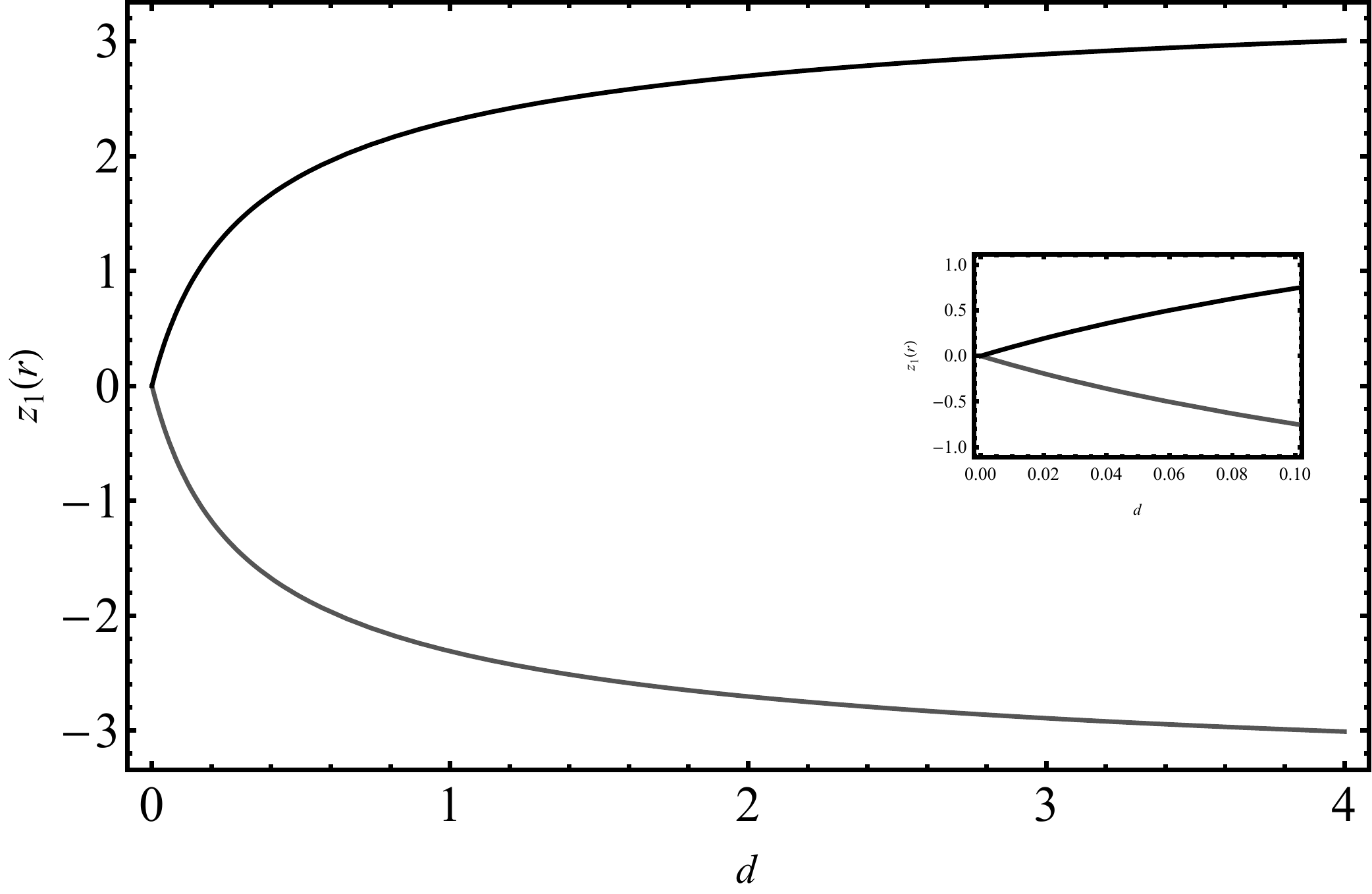}
    \includegraphics[width=0.9\columnwidth]{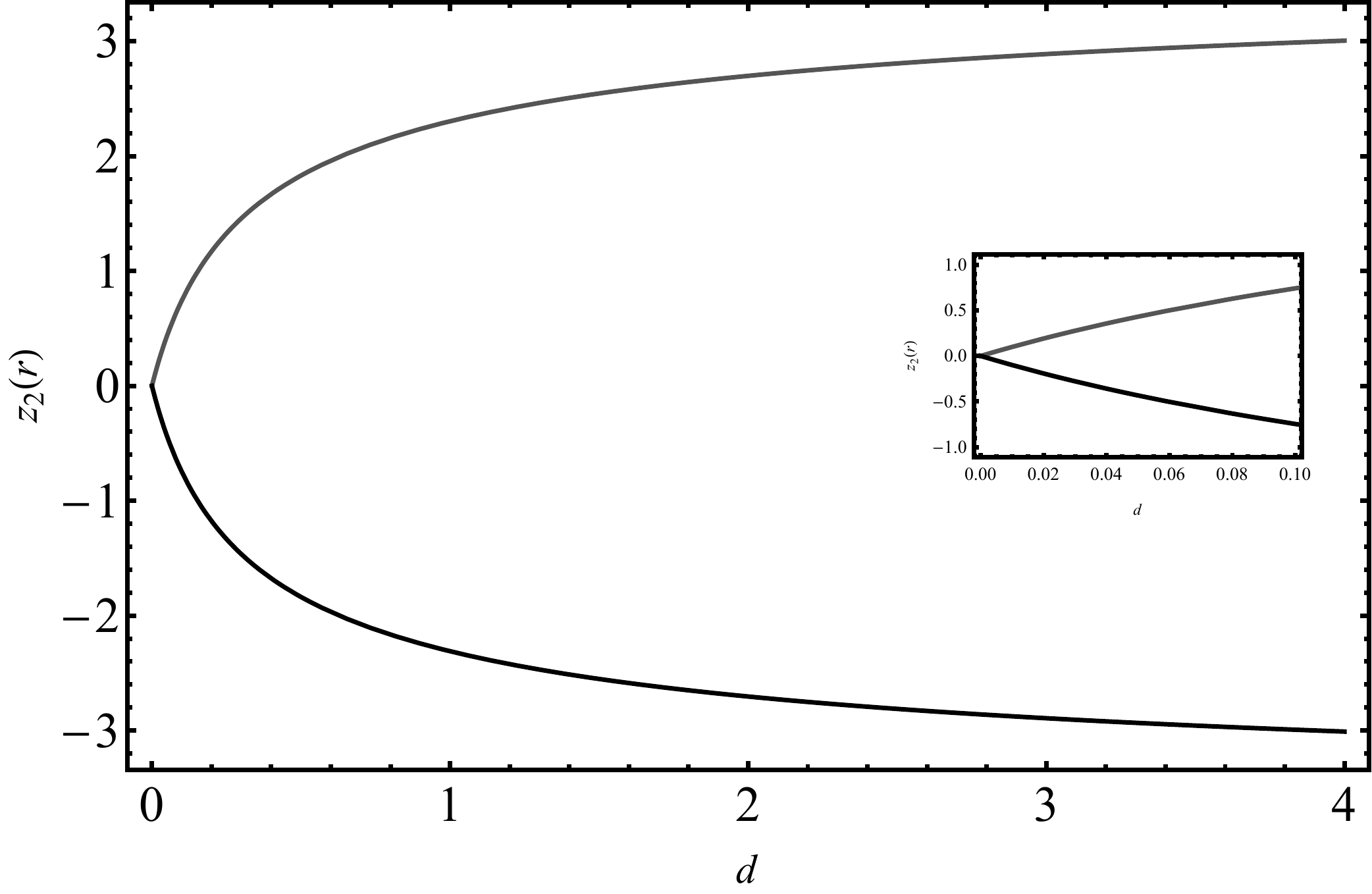}
    \caption{\emph{$z_1(r)$ and $z_2(r)$ as function of $d=r_d-r_e$ within the Schwarzschild - de Sitter spacetime. The gravitational field source is a white dwarf of mass $M=0.18M_\odot$, with $M_\odot=1.47$ km the solar mass, and radius $R=18304.5$ km, in the maximally - rotating configuration \cite{2.7}. Here,  $d\in[0;4\cdot 10^5]$ km, whereas  $z_1$ and $z_2$ are in power of $10^{-3}$. In the small zoom, we report  $z_1$ and $z_2$ up to $d=10^4$ km.} }
      \label{fig:WDSDS}
    \end{center}

\end{figure}

In the case of WD, we have regimes of intermediate gravity. We therefore compute our expectations over $z_1(r)$ and $z_2(r)$ in the maximally - rotating configuration, with the ranges of masses and radii respectively given by $M\in[0.18;1.47]M_\odot$ and $r\simeq 18304.5$ km. The values got by $z_1$ and $z_2$ reach a plateau as $d\gtrsim3$, {\it i.e.}, as $d$ becomes larger than zero. As well as NS regime for both the setups of Schwarzschild - de Sitter and Zipoy - Voorhees spacetimes, with very large $\delta$, we encounter the same behaviors. As for the NS, we can deduce that the Schwarzshild - de Sitter metric is even a good approximation  for  WDs in determining the red and the blue shift. Finally, from Figs. (\ref{fig:WDZV}) - (\ref{fig:WDSDS}) we immediately get
\begin{eqnarray}
    &z_{1+,\,{\rm ZV}}\,=\,z_{2-,\,{\rm ZV}}\,=\,z_{1-,\,{\rm SdS}}\,=\,z_{2+,\,{\rm SdS}}\,,\label{simmetrie1}\\
    &z_{1-,\,{\rm ZV}}\,=\,z_{2+,\,{\rm ZV}}\,=\,z_{1+,\,{\rm SdS}}\,=\,z_{2-,\,{\rm SdS}}\,,\label{simmetrie2}
\end{eqnarray}
where the subscripts ${\rm ZV}$ and ${\rm SdS}$ indicate Zipoy - Voorhees and Schwarzschild - de Sitter spacetimes respectively.

\subsection{Regime of low gravity}

The use of lunar laser ranging is commonly got using powerful pulsed searchlight, from the Earth to lunar corner cube retroreflectors\footnote{Device made up of three perpendicular reflective surfaces which retro-reflect the signal in the same direction in which it arrived.} \cite{LR1,LR2}. More generally, Laser Ranging (LR) is a technique that gets a measure of the round - trip time of a laser fired by a ground station on Earth, received by a cube corner retroreflector on a satellite and reflected back to the station. If the satellite is not the Moon, we refer to as satellite laser ranging \cite{LR3}. We follow here both the lunar and satellite LR approaches to face the low gravity regime.
\\For every LR experiment is performed, it is important to take into account the relative motion between the laser source and the retroreflector. This causes an angular deflection of the laser beam. In fact, in absence of a relative motion, the laser would exactly return back to the station. The angular deflection is also called velocity aberration (VA) and it is defined as
\begin{equation}
\label{VA}
    VA=\frac{2}{c}\left[\Delta v_{d}-v_{e}\cos{\phi}\right],
\end{equation}
where $\Delta v_{d}$ is the difference between the orbital velocity and rotational velocity at the equator of the satellite or planet on which is placed the retroreflector, $v_{e}$ the rotational velocity of the satellite or planet on which is placed the laser source at the equator and $\phi$ the laser source latitude. Consequently, the term in the square brackets in Eq. (\ref{VA}) represents the relative velocity between the involved two objects.
\noindent
We are going to use the VA and the laser wavelength to determine the red and blue shifts through  relativistic Doppler effect in the equatorial plane, {\it i.e.},
\begin{equation}
    \lambda'=\lambda\sqrt{\frac{1+VA/2}{1-VA/2}}
\end{equation}
where, instead of the ratio between the relative velocity and $c$, we used the VA through Eq. (\ref{VA}). Thus, one immediately computes the required $z$ through the standard formula $z=\frac{\lambda'-\lambda}{\lambda}$.
\noindent
We intend to study the theoretical models by considering Earth and Mars as sources of the gravitational field in the solar system. We want to compare the theoretical results with the experimental ones deriving from the LR missions. We thus analyze our numerical technique for three configurations,

\begin{itemize}
    \item Earth - Moon system. Here, we model the two astrophysical objects and involve the lunar LR.

\noindent

    \item Earth - satellite system. Here, the configuration is analogous to the standard lunar LR, but using an artificial satellite instead of the Moon. This second case is therefore similar to the previous one\footnote{We here refer to the LARES2 expectations. For us, LARES2 is the acronym of  Laser RElativity Satellite No. 2 and makes use of the Earth - satellite LR prerogative that we need. }.

    \item Mars - Phobos system. Here we propose \emph{ex novo} an  architecture to get feasible red and blue shift ranges. So, our computation provides an alternative view to LR, since it involves a laser source placed on Mars, considered the gravitational source modeled through our spacetime approach. The corner cube retroreflectors could be placed on  Phobos, i.e. a natural Mars satellite\footnote{This prerogative goes beyond the binary system Earth - Moon, with the advantage of being closer to each other, i.e. guaranteeing a more precise pulse measure.} A similar approach has been developed by \cite{phobos}, where the authors proposed a system constituted by Earth and Phobos. The procedure we here develop permits to extend previous results through the use of alternative technologies of LR.
\end{itemize}
\noindent
Our predicted values, got from the theoretical models under exam for these three configurations, are portrayed in Tabs. \ref{table2}-\ref{table3}. Here,  we again observe the same symmetries reported in Eqs. (\ref{simmetrie1},\ref{simmetrie2}).

\begin{table*}[htp]
\begin{tabular}{|c|c|c|c|}
\hline
\hline
        $z\pm\delta z$ & Lunar LR ($ 10^{-5}$)       & Satellite LR ($10^{-6}$) & Phobos LR ($10^{-6}$) \\
\hline\hline
  &  &  &  \\
  $z_{1+}\equiv z_{2-}$        & $2.298042$   & $7.36990$  &  $4.72646$  \\
  &  &  &  \\
  $\delta z_{1+}\equiv \delta z_{2-}$        & $\pm 0.000005$   & $\pm 0.01551$  &   $\pm 0.00046$  \\
  &  &    &\\
  \hline\hline
  & & & \\
  $z_{1-}\equiv z_{2+}$        & $-2.298057$   & $-7.37018$  &  $-4.72653$  \\
  &  &  &  \\
  $\delta z_{1-}\equiv \delta z_{2+}$        & $\pm 0.000005$   & $\pm 0.01551$  &   $\pm 0.00046$  \\
  &  &    &\\
\hline
\hline
\end{tabular}
\vspace{0.3cm}
\caption{\emph{Table of red and blue shift values, predicted by means of the Zipoy - Voorhees metric. We also report the corresponding error bars, namely $\delta z_{1,2\pm}$, evaluated by the standard logarithmic error propagation. Since for low gravity,  the predictions over $z$ do not significantly change by varying  $\delta$, with an accuracy smaller than one part over $10^6$, we arbitrarily select $\delta=0.5$. Here we consider Lunar LR for the system Earth - Moon, satellite LR for the system Earth - LARES2 \cite{lares} and finally Phobos LR for the proposed experiment that employs Mars and its satellite, Phobos.}}
\label{table2}
\end{table*}
\noindent
\begin{table*}[htp]
\begin{tabular}{|c|c|c|c|}
\hline
\hline
       $z\pm\delta z$  & Lunar LR  ($ 10^{-5}$)     & Satellite LR ($ 10^{-6}$) & Phobos LR ($ 10^{-6}$)\\
\hline\hline
  &  &  &  \\
  $z_{1+}\equiv z_{2-}$        & $-2.298057$   & $-7.37018$  &  $-4.72653$  \\
  &  &  &  \\
  $\delta z_{1+}\equiv \delta z_{2-}$        & $\pm 0.000005$   & $\pm 0.01551$  &   $\pm 0.00046$  \\
  &  &  &  \\
  \hline\hline
  & & &  \\
  $z_{1-}\equiv z_{2+}$        & $2.298057$   & $7.37018$  &  $4.72653$  \\
  &  &    &\\
  $\delta z_{1-}\equiv \delta z_{2+}$        & $\pm 0.000005$   & $\pm 0.01551$  &   $\pm 0.00046$  \\
  &  &    &\\
\hline
\hline
\end{tabular}
\vspace{0.3cm}
\caption{\emph{Table of red and blue shift values, predicted by means of the Schwarzschild - de Sitter metric. We also report the corresponding error bars, namely $\delta z_{1,2\pm}$, evaluated by the standard logarithmic error propagation. Here we consider Lunar LR for the system Earth - Moon, satellite LR for the system Earth - LARES2 \cite{lares} and finally Phobos LR for the proposed experiment that employs Mars and its satellite, Phobos.}}\label{table3}
\end{table*}
\noindent
On the other hand, the indirect measurements of $z$, obtained starting from the experimental data of the LR, are instead reported in Tabs.  \ref{table5}-\ref{table7} in Appendix \ref{AppB}. The slight differences between experimental and predicted outcomes are due to several facts. First, experimentally speaking, we are handling the VA technique only. Second, the accuracy can be refined adopting more than the configurations here investigated. Below, we summarize how to heal such issues adopting direct measure methods, by proposing novel experimental configurations.

\subsubsection{Designs of proposed experimental setups}

In view of the overall results, we summarize the following consequences of our treatments.

\begin{itemize}
    \item At high gravity regimes $\delta$ values, using axisymmetric spacetime, agree with the Schwarzschild - de Sitter prediction, making use of the Planck satellite bounds.
    \item As well as NS, the regimes of intermediate gravity, here investigated employing WD, behave in analogy.

\item At low gravity, in the solar system, our spacetime metrics are clearly unadequate to fix stringent limits over the free parameters as well as the red and blue shift intervals that slightly disagree with observations, see Appendix B.
\item Analogously LR cannot be used to get indirect measurements for $z$. Moreover, the results coming from the use of Schwarzschild metric, without taking care about the $\Lambda$ value, seem to agree with our scheme, indicating that $\Lambda$ is quite badly constrained within the solar system (see Tab. \ref{table4} in Appendix \ref{appendix}).
\end{itemize}
\noindent
From the above considerations, a direct measure of $z$ would be more predictive than other indirect treatments. So that, one can build up experimental setups based on genuine wavelength measures only.
\noindent
To do so, let us take the simplest configuration we could work with, based on the system Earth - Moon. We can send a laser pulse whose wavelength is known, as well as in the lunar LR technique. The detector is meant to measure the corresponding wavelength and then to get possible hints on  red or blue shifts.
\noindent
Alternatively, another possibility is offered by a rotating orbiter around Mars or more away planets, without excluding to take into account other configurations. The orbiter sends signals, whose wavelength is known. The detector, again placed on the given planet we are considering, gets the signals and provides a direct measure of the pulse shift in wavelength that is converted in red and blue shift. Last but not least, even if the astrophysical configurations of NS and WD could in principle be adopted for future missions to better get  $z$, the  experimental complexity to put on them instruments would be a great limitation for the experiment itself. On the other hand, cosmological red and blue shifts would be the key to fix the cosmological constant value. The strategy would be to take the large scale structure of the universe, switching the spacetime to more complicated metrics that could overcome the likely issues related to Schwarzschild - de Sitter.

\subsection{Error analysis}

In this subsection, we report the values of $z$ and the respective errors obtained when the field sources are the NS and the WD, both in the two employed metrics. We have selected $3$ values of $d = r_d - r_e$ within the range chosen for the plots, $d\in[0; 4\cdot 10^5]$. In particular:
\begin{itemize}
    \item $d = 0$ and $d = 4\cdot 10^5$, i.e. the extremes of the interval;
    \item $d = 2\cdot 10^5 $, which is the midpoint of the interval.
\end{itemize}
In the case of the Zipoy - Voorhees metric, we considered the $\delta$ values chosen for the plots ($\delta = 1/4$, $\delta = 3/4$ and $\delta = 1000$); while for the Schwarzschild - de Sitter metric, we considered the $\Lambda$ value of the Planck collaboration: $\Lambda=(1.10566\pm0.022703)\cdot 10^{-46}\;{\rm km^{-2}}$.
\\Errors are calculated through the standard logarithmic error propagation, considering that for NS and WD it is hard to get with high accuracy both mass and radii. We therefore work out the following strategy: we consider Refs. \cite{aggiunta1,aggiunta2} and there we got the maximum and minimum bounds associated to mass and radii for both NS and WD. Then, we consider the average constraints and use for our computation, in particular
\begin{itemize}
    \item for NS: $M=\left(1.07\pm0.11\right)M_\odot$ and $R=13.61^{+2.18}_{-0.68}$ km,
      \item for WD: $M=\left(0.180^{+0.056}_{-0.004}\right)M_\odot$ and $R=18304.5^{+5491.3}_{-823.70}$ km,
\end{itemize}
where $M_\odot=1.47$ km. Computation has been reported in Appendix \ref{errori}.

\section{Conclusions}\label{sezione5}

In this paper, we evaluated the red and blue shifts for distinct astrophysical and cosmological sources. To do so, we arbitrarily split three gravitational regimes, {\it i.e.}, high, intermediate and low gravity, based on the use of NS, WD and solar system constraints. We characterized two spacetimes, the first inspired by astrophysical configuration, exploring the consequences of the Zipoy - Voorhees metric on NS and WD. The second based on cosmological consideration, using the Schwarzschild - de Sitter spacetime. By varying the free parameters that enter the two metrics, we got feasible red and blue shift intervals and interpreted our expectations in view of current experiments and limits.
\noindent
Since the two underlying spacetimes are mostly different, we consider an axisymmetric solution, {\it i.e.}, the Zipoy - Voorhees metric, to characterize NS and WD, showing that at the level of solar system, the $\delta$ free term is unbounded. Analogously, at low gravity, we assumed, instead, the Schwarzschild - de Sitter solution, where we fixed the $\Lambda$ Planck's value, getting acceptable red and blue shift ranges. Further, we considered lunar and satellite LR techniques and showed an overall overestimation on $z_{1,2\pm}$ when using the VA. In turn, we interpreted such a result noticing that at low gravity the general relativistic effects are clearly disfavored. In fact, even the $\Lambda$ value  predicted by the Planck satellite cannot fully reproduce the expected intervals of $z_{1,2\pm}$ that LR  experiments estimate.
\noindent
On the other side, bearing in mind the maximally - rotating configurations for NS and WD, we got suitable red and blue shift intervals. We therefore concluded the most suitable approximations for NS and WD objects could be performed involving high quadrupole moments. The same happened for the cosmological constant, besides the solar system regime, showing a good agreement with current bounds and indicating the goodness of Planck $\Lambda$ measurements. Error bars have been computed for specific cases got from experiments over NS and WD mass and radii, in full agreement with our predicted bounds.
\noindent
Possible experimental designs for improving the quality of our results have been naively proposed. We discussed coarse - grained approaches to build up likely experimental configurations and set ups with the aim of refining the current accuracy over red and blue shift. To this end, we propose to adopt the binary system composed by Mars - Phobos to improve $z_{1,2\pm}$ measurements using the satellite LR technique.
\noindent
As red and blue shifts can be used to test the equivalence principle and/or sometimes to check the validity of particular classes of models, we intend to contrive new experiments that will be able to construct bounds over $\Lambda$, instead of postulating it. In so doing, as perspective we expect to work out a back - scattering procedure, different from the one here developed. Moreover, we intend to model other astrophysical sources as possible probes to test the red and blue shifts at different regimes of gravity, with the ambitious aim of reconciling several gravity regimes. In this respect, we underline that a more detailed analysis will be expected for getting experimental bounds in cosmology. So, the proposed experimental set up, with additional requirements that will be studied in incoming efforts, could represent a new technique to alleviate tensions between different measurements of $H_0$ in cosmology \cite{tensione}. We believe this can be also generalized for any other cosmological tension, in general.

\section*{Acknowledgments}
OL  acknowledges funds from the Ministry of Education and Science of the Republic of Kazakhstan, IRN  AP08052311.
LM acknowledges the support of Istituto Nazionale di Fisica Nucleare (INFN), iniziativa specifica MOONLIGHT2.

\appendix
\section{Results for the Schwarzschild spacetime}
\label{appendix}
The Schwarzschild metric reads
\begin{equation}
\label{Schwarzschild}
    ds^2=-\left(1-\frac{2m}{r}\right)dt^2+\frac{1}{\left(1-\frac{2m}{r}\right)}dr^2+r^2d\Omega^2,
\end{equation}
with $d\Omega^2=d\theta^2+\sin^2\theta d\varphi^2$. Below, we summarize the main results of the general procedure, described in Sec. \ref{sezione2}, applied to this metric. In particular, we find an overall agreement with the results got in   \cite{2.2}. We have as conserved quantities
\begin{eqnarray}
&&E=\left.\frac{r-2m}{\sqrt{r(r-3m)}}\right|_{r=r_p},\\
&&L=\left.\pm r\sqrt{\frac{m}{r-3m}}\right|_{r=r_p},
\end{eqnarray}

\noindent and velocities

\begin{eqnarray}
&&\left.u^t\right|_{r=r_p}=\left.\sqrt{\frac{r}{r-3m}}\right|_{r=r_p},\\
&&\left.u^{\varphi}\right|_{r=r_p}=\left.\pm\frac{1}{r}\sqrt{\frac{m}{r-3m}}\right|_{r=r_p},
\end{eqnarray}
and the kinematic quantities
\begin{eqnarray}
&& b_{\pm}=\mp \frac{r}{\sqrt{1-\frac{2m}{r}}},\\
&&\Omega_{d\pm}=\pm\sqrt{\frac{m}{r_d^3}}.
\end{eqnarray}

Finally, $z_{1\pm}$ and $z_{2\pm}$ are
\begin{eqnarray}
\label{z1_S}
z_{1\pm}&=&\pm\frac{\sqrt{\frac{r_e}{r_e-3m}}\left(\frac{1}{\sqrt{r_d-2m}}-\frac{1}{\sqrt{r_e-2m}}\right)}{\sqrt{\frac{r_d}{r_d-3m}}\left[1\mp\sqrt{\frac{m}{r_d-2m}}\right]},\\
\,\nonumber\\
z_{2\pm}&=&\mp\frac{\sqrt{\frac{r_e}{r_e-3m}}\left(\frac{1}{\sqrt{r_d-2m}}-\frac{1}{\sqrt{r_e-2m}}\right)}{\sqrt{\frac{r_d}{r_d-3m}}\left(1\pm\sqrt{\frac{m}{r_d-2m}}\right)}.\label{z2_S}
\end{eqnarray}

\begin{table}[htp]
\scriptsize
\begin{tabular}{|c|c|c|c|}
\hline
\hline
       $z\pm\delta z$  & Lunar LR  ($10^{-5}$)     & Satellite LR ($10^{-6}$) & Phobos LR ($10^{-6}$)\\
\hline\hline
  &  &  &  \\
  $z_{1+}\equiv z_{2-}$        & $-2.298057$   & $-7.37018$  &  $-4.72653$  \\
  &  &  &  \\
  $\delta z_{1+}\equiv \delta z_{2-}$        & $\pm 0.000005$   & $\pm 0.01551$  &   $\pm 0.00046$  \\
  &  &  &  \\
  \hline\hline
  & & &  \\
  $z_{1-}\equiv z_{2+}$        & $2.298057$   & $7.37018$  &  $4.72653$  \\
  &  &    &\\
  $\delta z_{1-}\equiv \delta z_{2+}$        & $\pm 0.000005$   & $\pm 0.01551$  &   $\pm 0.00046$  \\
  &  &    &\\
\hline
\hline
\end{tabular}
\vspace{0.3cm}
\caption{\emph{Table of red and blue shift values, predicted by means of the Schwarzschild metric. We also report the corresponding errors, namely $\delta z_{1,2\pm}$, evaluated by the standard logarithmic error propagation. Here we consider Lunar LR for the system Earth - Moon, satellite LR for the system Earth - LARES2 \cite{lares} and finally Phobos LR for the proposed experiment that employs Mars and its satellite, Phobos.}}\label{table4}
\end{table}
\noindent
Comparing these last two equations with those obtained for the Zipoy - Voorhees metric and for the Schwarzschild - de Sitter metric, we easily note that Eqs. (\ref{z1_ZV}) - (\ref{z2_ZV}) reduce to Eqs. (\ref{z1_S}) - (\ref{z2_S}) for $\delta=1$ (that is $k=m$) and Eqs. (\ref{z1_SdS}) - (\ref{z2_SdS}) reduce to Eqs. (\ref{z1_S}) - (\ref{z2_S}) for $\Lambda=0$.

\newpage
\section{Numerical constraints got from LR experiments}
\label{AppB}

In this appendix, we report a few tables concerning the LR experiments, emphasizing the main differences among lunar LR and satellite LR. Although different types of CCRs used in space missions, to calculate $z$ we only require the VAs depending on the orbital and rotational involved objects velocities, the local position of the laser source on the planet and the laser beam wavelength, generally $532$ nm.

\begin{table}[htp]
\begin{tabular}{|c|c|c|c|}
\hline
\hline
Station          & $\phi$       & VA $  (10^{-6})$& $z \pm\delta z$ $ (10^{-6})$   \\
\hline\hline
  &  &    &\\
  McDonald        & \ang{30}.68 N   & $4.051$  &   $2.025629 \pm 0.000007$  \\
  &  &    &\\
 Apollo       & \ang{32}.78 N   & $4.112$  &   $2.055989 \pm 0.000007$   \\
 & &  &\\
  \hline\hline
  & &  &\\
  Matera       & \ang{40}.65 N   & $4.370$  &   $2.185235 \pm 0.000007$    \\
  &  &   &\\
  Grasse       & \ang{43}.75 N   & $4.485$  &   $2.242475 \pm 0.000007$ \\
  & &  &\\
\hline
\hline
\end{tabular}
\vspace{0.3cm}
\caption{\emph{z value for four different operational LR stations placed on Earth and retroreflectors placed on the Moon. The velocities are expressed in units of radiant. We also report the corresponding errors, namely $\delta z$,  evaluated by the standard logarithmic error propagation. The table is split into $\phi<40$ and $\phi>40$. The increase of $z$ in function of $\phi$ is more than $10\%$ for largest $\phi$ here reported. The mean value is: $(2.127332\pm 0.000007)\cdot 10^{-6}$.}}
\label{table5}
\end{table}

\begin{table}[htp]
\begin{tabular}{|c|c|c|c|}
\hline
\hline
Station          & $\phi$       & VA $( 10^{-6})$& $z \pm \delta z$ $(10^{-5})$  \\
\hline\hline
  &  &  &  \\
  McDonald        & \ang{30}.68 N   & $35.30$  &   $1.76428 \pm 0.00159$  \\
  &  &   &\\
 Apollo       & \ang{32}.78 N   & $35.36$  &   $1.76826 \pm 0.00159$   \\
  & &  &\\
  \hline\hline
  & &  &\\
  Matera       & \ang{40}.65 N   & $35.62$  &   $1.78118 \pm 0.00159$   \\
  &  &    &\\
  Grasse       & \ang{43}.75 N   & $35.74$  &   $1.78690 \pm 0.00159$ \\
&  &    &\\
\hline
\hline
\end{tabular}
\vspace{0.3cm}
\caption{\emph{$z$ value for four different operational LR stations placed on Earth and retroreflectors placed on LARES2 satellite \cite{lares}. The velocities are expressed in units of radiant. We also report the corresponding errors, namely $\delta z$,  evaluated by the standard logarithmic error propagation. The table is split into $\phi<40$ and $\phi>40$. The increase of $z$ in function of $\phi$ is more than $10\%$ for largest $\phi$ here reported. The mean value is: $(1.775155\pm 0.00159)\cdot 10^{-5}$.}}
\label{table6}
\end{table}
\begin{table}[htp]
\begin{tabular}{|c|c|c|c|}
\hline
\hline
Station          & $\phi$       & VA $( 10^{-6})$& $z \pm\delta z$ $(10^{-6})$  \\
\hline\hline
  &  &    &\\
  Mars        & \ang{0}   & $12.63$  &   $6.314072 \pm 0.000052$  \\
  &  &    &\\
\hline
\hline
\end{tabular}
\vspace{0.3cm}
\caption{\emph{In this table, we report a plausible $z$ value for a station placed on Mars equator, with the ansatz that a few retroreflectors will be placed on Phobos, as we proposed above. We also report the corresponding errors, namely $\delta z$,  evaluated by the standard logarithmic error propagation. In particular, we figure out a laser station placed on Mars equator, {\it i.e.}, the position in which  satellite views got from Mars leads to the best configuration. }}
\label{table7}
\end{table}

\section*{Changing the plane}

For our spacetimes, the dependence on $\theta$ is contained only in the $g_{\varphi\varphi}$ component of the metric tensor, through $\sin^2{\theta}$. Moreover, these metrics are invariant under the transformation $
    \theta\rightarrow\theta'=\pi\pm\theta$
since $\sin^2{\theta'}=\sin^2{\theta}$. It is therefore intriguing to discuss below some consequences of this recipe.

\paragraph*{Schwarzschild - de Sitter metric. }

Let us fix the value of $\theta$ and, to be as general as possible, let us denote this value with $x$. Therefore, the metric tensor components useful for the calculation are
\begin{eqnarray}
g_{tt}&=&-\left(1-\frac{2m}{r}+\frac{\Lambda r^2}{3}\right),\\
g_{\varphi\varphi}&=&r^2\sin^2{x}.
\end{eqnarray}
Solving the system of equations
\begin{eqnarray}
    V_{{\rm eff}}(r_p)&=&0\\
    V'_{{\rm eff}}(r_p)&=&0
\end{eqnarray}
this time we get
\begin{eqnarray}
E&=&\frac{kr_p^3+3r_p-6m}{3\sqrt{r_p(3r_p-3m)}},\\
L&=&\pm\sin{(x)}r_p\sqrt{\frac{3m+kr_p^3}{r_p-3m}}.
\end{eqnarray}
Let us observe that the energy is independent from the value of $x$ (in fact, this is the same equation obtained in the case $\theta=\pi/2$), while the angular momentum depends on $x$, as we expected. Consequently, we have
\begin{eqnarray}
u^t\left|_{r_p}\right.&=&\sqrt{\frac{r_p}{r_p-3m}},\\
u^\varphi\left|_{r_p}\right.&=&\pm\frac{1}{\sin{x}}\sqrt{\frac{3m+kr_p^3}{3r_p^2(r_p-3m)}}
\end{eqnarray}
and
\begin{eqnarray}
b_{\pm}&=&\mp\sin{x}\frac{r}{\sqrt{1-\frac{2m}{r}+\frac{kr^2}{3}}},\\
\Omega_d&=&\pm\frac{1}{\sin{x}}\sqrt{\frac{3m+kr_d^3}{3r_d^3}}.
\end{eqnarray}
If we insert all these quantities\footnote{Let us remember that $c=e,d$.} into the following equations
\begin{equation}
    \label{z1}
    z_1=\frac{(u_e^t \Omega_d -u_e^{\varphi})b_{-}}{u_d^t(1-\Omega_d b_{-})},
\end{equation}
and
\begin{equation}
    \label{z2}
     z_2=\frac{(u_e^t \Omega_d -u_e^{\varphi})b_{+}}{u_d^t(1-\Omega_d b_{+})}
\end{equation}
we get that all the terms containing $\sin{x}$ simplify each other: the results for $z_1$ and $z_2$ are identical to those obtained with $\theta=\pi/2$. Therefore, without loss of generality, we can fix $\theta=\pi/2$ from the beginning.

\paragraph*{Zipoy - Voorhees metric.} Let us repeat the same analysis just done for the Schwarzschild - de Sitter metric. Thus, let us fix the value of $\theta$ and, to be as general as possible, let us denote this value with $x$. In this case, the metric tensor components useful for the calculation are
\begin{eqnarray}
g_{tt}&=&-\left(1-\frac{2k}{r}\right)^{\delta},\\
g_{\varphi\varphi}&=&\frac{r^2-2kr}{\left(1-\frac{2k}{r}\right)^{\delta}}\sin^2{x}.
\end{eqnarray}
Solving the system of equations
\begin{eqnarray}
    V_{{\rm eff}}(r_p)&=&0\\
    V'_{{\rm eff}}(r_p)&=&0
\end{eqnarray}
now we get
\begin{equation}
\label{E_ZV}
E=\sqrt{\frac{\left(1-\frac{2 k}{r_p}\right)^{\delta } (\delta  k+k-r_p)}{2 \delta  k+k-r_p}},
\end{equation}
\begin{equation}
\label{L_ZV}
L=\pm\sin{x}\sqrt{\frac{\delta  k r_p (2 k-r_p) \left(1-\frac{2 k}{r_p}\right)^{-\delta }}{2 \delta  k+k-r_p}}.
\end{equation}
Also for the Zipoy - Voorhees metric, the energy is independent from the value of $x$ (in fact, this is the same equation obtained in the case $\theta=\pi/2$), while the angular momentum depends on $x$, as we expected. Consequently, we have
\begin{equation}
\label{utZV}
\left.u^t\right|_{r=r_p}=-\sqrt{\frac{\left(1-\frac{2 k}{r_p}\right)^{-\delta } (\delta  k+k-r_p)}{2 \delta  k+k-r_p}},
\end{equation}
\begin{equation}
\label{ufZV}
\left.u^{\varphi}\right|_{r=r_p}=\pm\frac{1}{\sin{x}} \sqrt{\frac{\delta  k (2 k-r_p) \left(1-\frac{2 k}{r_p}\right)^{\delta }}{r_p (k-r_p)^2 (2 \delta  k+k-r_p)}}
\end{equation}
and
\begin{eqnarray}
 b_{\pm}&=&\mp\sin{x}\frac{\sqrt{r^2-2kr}}{\left(1-\frac{2 k}{r}\right)^{\delta}},\\
\Omega_{d\pm}&=&\mp\frac{1}{\sin{x}}\sqrt{\frac{\delta  k (2 k-r_d) \left(1-\frac{2 k}{r_d}\right)^{2 \delta }}{r_d (k-r_d)^2 (\delta  k+k-r_d)}}.
\end{eqnarray}
If we insert all these quantities into Eqs. (\ref{z1})-(\ref{z2}), we get again that all the terms containing $\sin{x}$ simplify each other: the results for $z_1$ and $z_2$ are identical to those obtained with $\theta=\pi/2$. Therefore, also for the Zipoy - Voorhees metric, without loss of generality, we can fix $\theta=\pi/2$ from the beginning.

\section*{Error propagation}\label{errori}

We report here the numerical values evaluated from our error propagation procedure, making use of the logarithmic error propagation rule.

\begin{table*}[htp]
\centering
\caption{\footnotesize{z values with errors for the Zipoy - Voorhees metric, with a NS and a WD as field sources, for $3$ different values of $\delta$, at $3$ different values of $d$.}}
\begin{tabular}{|c|c|c|c|c|c|c|}
\hline
                           & \multicolumn{2}{c|}{$d=0$} & \multicolumn{2}{c|}{$d=2\cdot 10^5$} & \multicolumn{2}{c|}{$d=4\cdot 10^5$}  \\
\hline
$\delta=1/4$  & NS & WD                  & NS & WD                          & NS & WD                           \\
\hline
                           &    &                     &    &                             &    &                              \\
$z_{1+}=z_{2-}$        & $0.00000^{+0.11185}_{-0.03489}$  & $0.00000^{+0.00057}_{-0.00009}$                    &$0.11412^{+0.25949}_{-0.15085}$    & $0.00265^{+0.00098}_{-0.00011}$                            & $0.11555^{+0.25894}_{-0.15050}$    & $0.00299^{+0.00103}_{-0.00012}$                             \\
                           &    &                     &    &                             &    &                              \\
                           &    &                     &    &                             &    &                              \\
$z_{1-}=z_{2+}$        & $0.00000^{+0.12958}_{-0.04042}$  & $0.00000^{+0.00057}_{-0.00009}$                    &$-0.11477^{+0.26092}_{-0.15166}$    &  $-0.00265^{+0.00098}_{-0.00012}$                            &  $-0.11601^{+0.25995}_{-0.15107}$   & $-0.00299^{+0.00104}_{-0.00012}$                              \\
                           &    &                     &    &                             &    &                              \\
\hline
$\delta=3/4$  & NS & WD                  & NS & WD                          & NS & WD                           \\
\hline
                           &    &                     &    &                             &    &                              \\
$z_{1+}=z_{2-}$        & $0.00000^{+0.01657}_{-0.00517}$  & $0.00000^{+0.00057}_{-0.00009}$                    & $0.40249^{+0.07605}_{-0.04408}$    &  $0.00265^{+0.00098}_{-0.00011}$                           & $0.40385^{+0.07626}_{-0.04423}$   &    $0.00299^{+0.00103}_{-0.00012}$                          \\
                           &    &                     &    &                             &    &                              \\
                           &    &                     &    &                             &    &                              \\
$z_{1-}=z_{2+}$        & $0.00000^{+0.03254}_{-0.01015}$  & $0.00000^{+0.00057}_{-0.00009}$                    & $-0.40475^{+0.07659}_{-0.04444}$   &    $-0.00265^{+0.00098}_{-0.00012}$                         & $-0.40545^{+0.76646}_{-0.04449}$   &   $-0.00299^{+0.01037}_{-0.00012}$                          \\
                           &    &                     &    &                             &    &                              \\
\hline
$\delta=1000$ & NS & WD                  & NS & WD                          & NS & WD                           \\
\hline
                           &    &                     &    &                             &    &                              \\
$z_{1+}=z_{2-}$        & $0.00000^{+0.02404}_{-0.00750}$  & $0.00000^{+0.00057}_{-0.00009}$                    &  $0.43056^{+0.08699}_{-0.05044}$  &    $0.00265^{+0.00098}_{-0.00011}$                         &  $0.43194^{+0.08719}_{-0.05058}$  &  $0.00299^{+0.00103}_{-0.00012}$                            \\
                           &    &                     &    &                             &    &                              \\
                           &    &                     &    &                             &    &                              \\
$z_{1-}=z_{2+}$       & $0.00000^{+0.05126}_{-0.01599}$  & $0.00000^{+0.00057}_{-0.00009}$                   & $-0.43302^{+0.08761}_{-0.05085}$   &  $-0.00265^{+0.00984}_{-0.00012}$                           & $-0.43366^{+0.08762}_{-0.05087}$   & $-0.00299^{+0.00104}_{-0.00012}$                             \\
                           &    &                     &    &                             &    &                              \\
\hline
\end{tabular}
\end{table*}

\begin{table*}[htp]
\centering
\caption{\footnotesize{z values with errors for the Schwarzschild - de Sitter metric, with a NS and a WD as field sources, at $3$ different values of $d$.}}
\begin{tabular}{|c|c|c|c|c|c|c|}
\hline
                    & \multicolumn{2}{c|}{$d=0$} & \multicolumn{2}{c|}{$d=2\cdot 10^5$} & \multicolumn{2}{c|}{$d=4\cdot 10^5$}  \\
\hline
                    & NS & WD                  & NS & WD                          & NS & WD                           \\
\hline
                    &    &                     &    &                             &    &                              \\
$z_{1+}=z_{2-}$ & $0.00000^{+0.06596}_{-0.02057}$  & $0.00000^{+0.00057}_{-0.00009}$                    & $-0.47754^{+0.11548}_{-0.06703}$   &    $-0.00266^{+0.00098}_{-0.00012}$                         &  $-0.47816^{+0.11549}_{-0.06705}$  & $-0.00299^{+0.00104}_{-0.00012}$                             \\
                    &    &                     &    &                             &    &                              \\
                    &    &                     &    &                             &    &                              \\
$z_{1-}=z_{2+}$ & $0.00000^{+0.06596}_{-0.02057}$  & $0.00000^{+0.00057}_{-0.00009}$                    & $0.47754^{+0.11578}_{-0.06703}$   &  $0.00266^{+0.00098}_{-0.00012}$                            &  $0.47816^{+0.11549}_{-0.06705}$   &  $0.00299^{+0.00104}_{-0.00012}$                             \\
                    &    &                     &    &                             &    &                              \\
\hline
\end{tabular}
\end{table*}
\end{document}